\documentclass[a4paper,fleqn]{cas-sc}

\usepackage{todonotes}
\usepackage{arydshln}
\usepackage[utf8]{inputenc}
\usepackage{multirow}
\usepackage{makecell}
\usepackage{caption}
\usepackage{afterpage}
\usepackage[style=apa, backend=biber]{biblatex}

\def\tsc#1{\csdef{#1}{\textsc{\lowercase{#1}}\xspace}}
\tsc{WGM}
\tsc{QE}

\makeatletter                    
\newlength{\bibsep}{\@listi \global\bibsep\itemsep \global\advance\bibsep by\parsep}
\makeatother  

\begin{document}
\let\WriteBookmarks\relax
\def\floatpagepagefraction{1}
\def\textpagefraction{.001}

\shorttitle{Revealing complexities when adult readers engage in the credibility evaluation of social media posts}    

\shortauthors{<short author list for running head>}  

\title [mode = title]{Revealing complexities when adult readers engage in the credibility evaluation of social media posts}  

\cortext[1]{Corresponding author}

\affiliation[1]{organization={Faculty of Computer Science, Dalhousie University},
            addressline={6050 University Avenue, PO BOX 15000}, 
            city={Halifax, N S},
            postcode={B3H 4R2}, 
           country={Canada}}

\affiliation[2]{organization={Faculty of Information Technology and Electrical Engineering, University of Oulu},
            addressline={P.O. Box 4500}, 
            city={Oulu},
            postcode={90014}, 
           country={Finland}}
\affiliation[2]{organization={Faculty of Education and Culture, Tampere University},
            addressline={P.O. Box 700}, 
            city={Tampere},
            postcode={33014}, 
           country={Finland}}

\affiliation[4]{organization={Department of Education and Lifelong Learning, Faculty of Social and Educational Sciences, Norwegian University of Science and Technology},
            addressline={P.O. Box
8900, Torgarden NO-7491}, 
            city={Trondheim},
           country={Norway}}

  \affiliation[5]{organization={Tokyo College, University of Tokyo},
      addressline={1 Chome-1-1, Bunkyo City, 113-0032},
       country={Japan}}

  \affiliation[6]{organization={Department of Computer Science, University of Helsinki},
      addressline={PL 68}, 
       city={Helsinki},
       postcode={00014}, 
       country={Finland}}

\affiliation[7]{organization={Feinstein College of Education, University of Rhode Island},
      addressline={615 Chafee Hall, 142 Flagg Road}, 
       city={Kingston},
       postcode={RI 02881},
       country={United States}}

\author[1,2]{Miikka Kuutila}[
      orcid=0000-0002-3695-7280]
\cormark[1]
\author[3]{Carita Kiili}[
      orcid=0000-0001-9189-4094]
\author[3,4]{Reijo Kupiainen}[
      orcid=0000-0003-2610-2294]
\author[2]{Eetu Huusko}[
      orcid=0000-0002-8882-8501]
\author[2]{Junhao Li}[
      orcid=0000-0002-0982-0158]
\author[2, 5]{Simo Hosio}[
      orcid=0000-0002-9609-0965]
\author[2,6]{Mika M{\"a}ntyl{\"a}}[
      orcid=0000-0002-2841-5879]
\author[7]{Julie Coiro}[
      orcid=0000-0002-2578-3748]
\author[3]{Kristian Kiili}[
      orcid=0000-0003-2838-6892]



\begin{abstract}
The internet, including social networking sites, has become a major source of health information for laypersons. Yet, the internet has also become a platform for spreading misinformation that challenges adults’ ability to critically evaluate the credibility of health messages. To better understand the factors affecting credibility judgements, the present study investigates the role of source characteristics, evidence quality, crowdsourcing platform, and prior beliefs of the topic in adult readers’ credibility evaluations of short health-related social media posts. Researchers designed content for the posts concerning five health topics by manipulating source characteristics (source’s expertise, gender, and ethnicity), accuracy of the claims, and evidence quality (research evidence, testimony, consensus, and personal experience) in the posts. Then, accurate and inaccurate posts varying in these other manipulated aspects were computer-generated. Crowdworkers (N = 844) recruited from two platforms were asked to evaluate the credibility of ten social media posts, resulting in 8380 evaluations. Before credibility evaluation, participants’ prior beliefs on the topics of the posts were assessed. Results showed that prior belief consistency and source expertise most affected the perceived credibility of accurate and inaccurate social media posts after controlling for the topic of the post. In contrast, the quality of evidence supporting the health claim mattered relatively little. In addition, the data collection platform had a notable impact, such that posts containing inaccurate claims were much more likely to be rated higher on one platform compared to the other. Implications for credibility evaluation theory and research are discussed.
\end{abstract}




\maketitle

\section{Introduction}

 The internet has become a major source of health information for laypersons \autocite{sun2019consumer}. Because information online can be published without adhering to verification standards \autocite{braasch2014sensitivity}, the internet has also become a platform for spreading misinformation, including issues concerning health \autocite{ecker2022psychological}. This requires a level of criticality that can challenge laypersons, especially when their prior knowledge about the issue at hand is limited \autocite{andreassen2012reading, braasch2014sensitivity}.
 
The challenge of credibility evaluation is even more pronounced when people encounter health-related issues on social networking sites. Social media messages are often very short and informal and may use improper syntax or spelling \autocite{addawood2016your}. On social networking sites, such as Twitter, short texts can contain arguments with accurate, inaccurate, inappropriate, or missing justifications, especially when these texts involve controversial topics. Although modern societies are portrayed by the division of cognitive labor wherein individuals defer to relevant experts to ground their understanding of less familiar topics \autocite{scharrer2017science}, on social networking sites anyone can share their knowledge, views, or opinions, regardless of their expertise.

Recent frameworks for credibility evaluation \autocites{barzilai2020dealing, forzani2022does} have depicted factors that readers can attend to when evaluating online information, including the quality of reasoning, consistency of content with one's prior knowledge or other relevant resources, the author's expertise and intentions, and text genre. Previous studies have mainly focused on a limited number of credibility evaluation aspects or employed experimental designs in which participants are assigned to specific conditions varying in credibility features. For example, \textcite{chinn2021effects} assigned participants to four conditions that differed in expertise (expert vs. layperson) and evidence type (statistical evidence vs. testimony) and then asked them to evaluate statements presented in the news articles. This study revealed an interaction between source expertise and evidence type such that the quality of evidence affected the perceived credibility of the laypersons but not that of the experts. 

However, studies designed to only manipulate features of content and source cannot reveal all the complexities involved in credibility evaluation.  Although there is substantial evidence that indicates variations in readers, texts, activities, and contexts can influence performance when making sense of information on the Internet \autocite{coiro2021toward}, few studies have investigated a more complex combination of factors (i.e., content, source, and context) that may predict how individuals evaluate the quality of social media texts and corresponding sources. Further, drawing research-based conclusions from crowdsourcing platforms whose membership represents mostly Western populations can lead to an imbalance of cultural and social perspectives \autocite{venturebeat2022how}. Consequently, sampling across platforms with more diverse membership can help examine whether the findings gleaned from Western populations hold in other regions around the globe. 

Therefore, the present study approached credibility evaluation more holistically by considering multiple factors, including the content (prior-belief consistency, evidence type), the source (expertise, gender, ethnicity) and the context (two crowdsourcing platforms representing multiple nationalities) across four health topics when adult readers evaluated either accurate or inaccurate social media posts. By choosing this approach, the study provides insights about the most dominating factors to consider when advancing theoretical frameworks and models of credibility evaluation. Moreover, this study advances the understanding of credibility judgments among the adult population as readers interact with information in globally networked social media contexts.

\section{Theoretical Model for First- and Second-Hand Evaluation }\label{sec:background}

Our study was primarily informed by a bidirectional model of first- and second-hand evaluation strategies \autocite{barzilai2020dealing} while also considering one additional feature of the Critical Online Resource Evaluation (CORE) framework \autocite{forzani2022does}. The bidirectional model of first- and second-hand evaluation strategies is based on the content-source-integration model \autocite{stadtler2014content} that suggests that readers employ two levels of strategies to resolve scientific conflicts. Readers can evaluate information validity to determine whether the presented claims are accurate (i.e., using first-hand evaluation). However, readers do not always have sufficient knowledge to make the appropriate judgment. In this case, they can resolve the conflict by turning to second-hand evaluation to determine whether the source of information is trustworthy. 

In their bidirectional model of first- and second-hand evaluation strategies, \autocite{barzilai2020dealing} specify three first-hand evaluation strategies to judge the validity of scientific claims (i.e., distinguish accurate claims from inaccurate ones). These strategies are knowledge-based validation, discourse-based validation, and corroboration. When using knowledge-based validation, readers rely on their prior knowledge and beliefs about the topic, whereas when using discourse-based validation, readers use various discourse features, such as argument structure and quality to make judgements about the credibility of content. Corroboration, a third strategy for validating content, is when readers compare text content to what other resources state about a given issue. 

The term second-hand evaluation \autocites{barzilai2020dealing, stadtler2014content} refers to sourcing strategies that readers employ to judge a source's trustworthiness. Readers can make inferences and evaluate a source’s expertise, benevolence, and integrity \autocite{hendriks2015measuring} by relying on available information about the source. According to the model, readers use first- and second-hand evaluation strategies reciprocally. That is, content quality judgments inform source trustworthiness judgments and vice versa. In addition to the first and second-hand evaluation strategies, readers can also evaluate context as part of the credibility evaluation process, which \textcite{forzani2022does} refers to as tertiary evaluation. According to \textcite{forzani2022does}, context refers to the temporal, social, and political setting of the text (e.g., social media texts read in specific crowdsourcing platforms). In this study, we considered the effects of platform context, while also incorporating a slightly broader interpretation of context by considering contextual attributes of the readers evaluating the texts (i.e., adult readers representing different nationalities).   

\subsection{First-Hand Evaluation} 
In this study, we focused on credibility judgments of health-related social media posts to determine the extent to which readers rely on their prior beliefs (i.e., knowledge-based validation) and information presented in the posts (i.e., discourse-based validation) and sourcing. Next, we define relevant concepts and outline previous research associated with first- and second-hand evaluation; first-hand evaluation involves knowledge-based and discourse-based validation while second-hand evaluation focuses on sourcing.  We also present research suggesting contextual attributes, such as a reader’s cultural background, can affect adults’ credibility evaluations in social media contexts.  

\subsubsection{Knowledge-Based Validation}

Prior knowledge and beliefs are internal resources that readers can use to quickly evaluate the accuracy of information \autocite{richter2017comprehension}. When readers evaluate information using their prior beliefs, they seem to favor belief-consistent information over belief-inconsistent information \autocites{abendroth2020mere, mccrudden2016differences, van2016attitude}. As such, readers attend to evidence that is consistent with their beliefs and resist evidence that contradicts their beliefs. This text-belief consistency effect is similar to confirmation bias \autocite{karimi2023thinking} - the term used in the psychological literature to refer to seeking or interpreting evidence in accordance with a person's existing beliefs, expectations, or hypotheses \autocite{nickerson1998confirmation}. However, a confirmation bias occurs when a reader starts with a particular view of a particular issue and then actively searches for additional information that upholds that view.  In contrast, the text-belief consistency effect occurs during the comprehension process \autocite{karimi2023thinking}, when a reader encounters information and, often unconsciously, attends to text that is consistent with their beliefs while also passing by text that contradicts their beliefs.

The text-belief consistency effect in credibility evaluation has been shown in different contexts, including isolated arguments with no indicators of context \autocite{mccrudden2016differences}, and arguments presented in the context of social media posts \autocite{wertgen2023source}. For example, \textcite{mccrudden2016differences} asked 72 high school students to evaluate eight isolated arguments about climate change. Four arguments claimed that humans affect climate change, and four other arguments claimed the opposite. Half of the arguments were strong, and half were weak. It was found that students evaluated belief-consistent arguments more favorably than belief-inconsistent arguments. However, arguments were not evaluated solely based on their belief-consistency. Namely, students evaluated strong arguments more favorably than weak arguments. Importantly, further analysis showed that students’ evaluations of strong belief-inconsistent arguments and weak belief-consistent arguments did not differ. 

In a second study, \textcite{wertgen2023source} examined the text-belief consistency effect when asking university students to evaluate short messages, resembling Twitter posts, on four socio-scientific controversial topics (see Experiment 1). For each topic, the messages represented opposing positions. In addition to the position, \textcite{wertgen2023source} also manipulated the source-message consistency. In the consistent source-message condition, the source (e.g., World Wildlife Fund) presented the expected position (humans cause climate change). In the inconsistent source-message condition, the source presented an unexpected position (natural causes for climate change). Participants read belief-consistent messages faster and judged them more plausible than belief-inconsistent messages. The same pattern was found for source-message consistency. When the sources presented expected positions in the social media messages, the messages were read faster and rated slightly more plausible (or more likely to be acceptable in a certain situation); when the sources presented unexpected  positions, the messages were read more slowly and rated slightly less plausible. Overall, the use of prior beliefs as a resource could be especially applicable when people initially browse short social media messages using quick, superficial processing \autocite{metzger2013credibility} before they stop to evaluate the information more strategically.  

\subsubsection{Discourse-Based Validation}
According to \textcite{barzilai2020dealing}, evaluating arguments presented in a text is an essential component of discourse-based validation. Discursive elements include the argument’s structure and coherence as well as the quality of evidence provided. Arguments may have a rhetorical structure or a dialogic structure. A rhetorical argument includes a claim (or assertion) with an accompanying justification, which differs from a dialogic argument or debate, which includes an opposition between two or more assertions \autocite{kuhn1991skills}. A simple rhetorical argument contains one claim with only one instance of justifying support, which is called evidence \autocite{zarefsky2019practice} or data \autocite{toulmin2003uses}. According to \textcite{toulmin2003uses}, an argument can also include other components, such as a warrant, which is an often unstated but implicit inference that connects the evidence to the claim. For example, empirical research (used as evidence) entitles us to conclude that vaccines are safe (the claim) because we trust that the research has followed scientific standards (the warrant). Social media posts with restricted lengths, such as those found on Twitter, are typically structured as simple rhetorical arguments, making them a relevant context for examining how adults attend to the coherence of a single claim, evidence, and warrant to evaluate the credibility of a simple argument.

In addition to judging the coherence of the parts of an argument, readers can also attend to the quality of evidence in an argument. There are several different types of evidence \autocites{HornikxJ.M.A2005Aroe, kuhn1991skills, zarefsky2019practice}; these types reflect how the evidence was produced and they vary in terms of their quality \autocite{ClarkA.Chinn2014ECaE}. Four types of evidence include evidence produced from research, testimonies, anecdotes, and social consensus \autocite{zarefsky2019practice}. Evidence resulting from rigorous research processes, such as experiments, statistical analyses, and meta-analyses, can offer causal and statistical evidence. A second type of evidence is testimonial evidence \autocite{zarefsky2019practice} or expert evidence \autocite{HornikxJ.M.A2005Aroe}. Testimony, which can consist of either a fact or an opinion, is a statement made by some source.  It is reasonable for an author to rely on testimonial evidence from a qualified source to support one’s claim when an author does not have direct knowledge of the topic but can trust the expertise of others. 

Anecdotal evidence, such as personal experience, relies on a narrative way of knowing \autocite{hinyard2007using} while social consensus refers to widely agreed-upon facts, shared value judgments, shared historical understandings, and previously established claims \autocite{zarefsky2019practice}. Anecdotal evidence and social consensus, two types of evidence that are not necessarily supported by accurate information (e.g., research evidence or expert testimony), are common in health communication, especially in social media contexts \autocite{suarez2021prevalence}. Thus, social media posts are an authentic context for determining if adults can evaluate the extent to which evidence supporting certain claims provides accurate and inaccurate information.

Because social media contain an enormous number of short messages competing for readers’ attention, the persuasiveness of evidence is crucial for authors to craft their arguments in ways that are both efficient and effective. When examining the relative persuasiveness of different evidence types (e.g., anecdotal versus statistical evidence), results have been contradictory. In one narrative literature review, \textcite{HornikxJ.M.A2005Aroe} reviewed twelve studies that compared the persuasiveness of anecdotal evidence and statistical evidence. Six of those studies showed that statistical evidence was more persuasive than anecdotal evidence, one study showed that anecdotal evidence was more persuasive, and five studies showed no difference between the two types of evidence.  Elsewhere, a meta-analysis of 15 studies concerning health communication campaigns \autocite{zebregs2015differential} showed that statistical evidence had a stronger influence on beliefs and attitudes than anecdotal evidence, but that anecdotal evidence had a stronger influence on intention, which relates more closely to affective responses. Notably, a recent review found that presenting anecdotal evidence and personal experiences was one of the 12 identified persuasive techniques used in communicating online health misinformation \autocite{peng2022persuasive}.

When evaluating evidence, readers may struggle to differentiate evidence in terms of quality, especially when judging the quality of different types of research evidence.  For example, \textcite{list2021examining} found that even though undergraduate students (\textit{N} = 82) considered anecdotal evidence considerably less convincing than other types of evidence, they struggled to differentiate between certain type of research evidence that imply causation and observational and correlational evidence (that do not imply causation). This is especially concerning because, on social media, discussions of health issues often include causal claims or generalizations that may or may not be supported by appropriate causal evidence.

\subsection{Second-Hand Evaluation}
The trustworthiness of a source refers to readers' perceptions of positive characteristics on the part of that source; these perceptions, allow them to accept the source’s message \autocites{fogg2002persuasive, ohanian1990construction}. Evaluation of source trustworthiness is essential, particularly when readers evaluate scientific information, including health information,  about which they do not have specialized knowledge \autocite{hendriks2015measuring}. \textcite{hendriks2015measuring} found three dimensions that laypersons rely on when deciding whether they trust an expert. These dimensions are expertise, integrity, and benevolence. A source’s \textit{expertise} can be evaluated by exploring that source’s credentials, such as degrees, professional achievements, affiliations, and other indicators of competence, such as awards and publications \autocites{rolin2020trust, braaten2018task}. Furthermore, \textit{integrity} refers to honesty, objectivity, and acting according to scientific and professional standards, whereas \textit{benevolence} refers to goodwill and intentions toward others and society \autocite{hendriks2015measuring}.		

It is widely acknowledged that information about a source's expertise plays a pivotal role in credibility evaluation of social media messages and beyond \autocites{meinert2022expertise, LIN2016264, stadtler2014content}. For example, \textcite{hirvonen2018cognitive} have studied cognitive authority in online health-related information-sharing processes. They found that authority-related cues, such as user information, the authority's own experience, education, and background, were important when girls and young women evaluated the information credibility of their peers in an online forum related to health topics. However, evaluation of the source's trustworthiness can also be based on symbols of authority and fruits of success rather than on perceived expertise \autocite{shieber2015testimony}.
For example, in one interview study, all 18 interviewed participants (ages 18 to 30) viewed celebrities as trustworthy sources of online information \autocite{djafarova2017exploring}. 

In other research, \textcite{LIN2016264} studied the credibility evaluation of mock Twitter posts that discussed health risks. In the study, 696 undergraduates participated in a quasi-experiment in which the following aspects of Twitter posts were manipulated: author information, originality of the tweets (original or retweeted), and bandwagon cues (i.e., replies or not). The authors were manipulated by authority and whether the identifying information was provided. The Twitter pages were owned by either the Center for Disease Control and Preventions (CDC), a college student, or a stranger without any identifying information. After viewing the pages, students were asked to evaluate the source's credibility according to three criteria: competence, goodwill, and trustworthiness. The researchers determined that participants viewed authority cues as the most credible compared to the other cues (i.e., identity, original or retweeted, bandwagon cues). The CDC was evaluated as more credible than the student and the stranger. The bandwagon cues also had an effect, such that the student and the stranger were evaluated as more credible when they retweeted each other. Interestingly, the CDC twitter page without any retweets was perceived as the most credible in all evaluated aspects.

Some studies have examined other features of the source, such as gender and ethnicity. \textcite{spence2013intercultural} investigated intercultural differences in responding to health messages specifically in the context of Facebook. In their study, participants from the Caucasian and African American communities (\textit{N} = 200\textit{)} were evaluated to determine their response efficacy and behavioral intentions after being exposed to a social media page with an avatar. The African American page owner communicated either high or low ethnic identity (that is, including a greater or lesser number of cues on their page to promote their ethnic identity) while encouraging participants to read a story about heart disease. African American participants perceived both ethnic authors as more competent, caring, and trustworthy than Caucasian participants did. While the perceived credibility of the author was not measured, African American participants indicated stronger intention to change their dietary behavior after seeing a health message delivered by an African American avatar with high ethnic identification as compared to one with low ethnic identification.

\textcite{armstrong2009blogs} examined the role of gender in credibility evaluation of blogs. They found that gender influenced how undergraduates at a large university in southeastern United States perceived credibility. The blog posts with male authors were seen as more credible than posts with female authors. Similar results regarding credibility and gender have also been attained with newscasters \autocite{weibel2008gender}. More recently, \textcite{groggel2019race} investigated how people perceived both gender and ethnicity as well as physical attractiveness as a cue indicating trustworthiness when evaluating Twitter accounts. In the study, Amazon Mechanical Turk workers evaluated the trustworthiness of 816 Twitter profiles. Results indicated that attractiveness was positively associated with trust. Furthermore, Twitter accounts belonging to white individuals were evaluated as more trustworthy than accounts belonging to Black individuals. Also, accounts belonging to Black males and both Black and White females were viewed as less trustworthy than accounts belonging to White males. White crowdsourcing workers also evaluated accounts belonging to White individuals as more trustworthy.

\subsection{Contextual Attributes of Readers}
At least two studies have found that cultural context or a reader's membership in a particular social media network influenced adults’ credibility perceptions of social media posts. \textcite{yang2013microblog} compared credibility perceptions of microblog posts among U.S. and Chinese participants in two social media networks; the posts varied in terms of author’s gender, name style, profile image, location, and degree of network overlap with the reader. They identified several key differences in how users from each country critically consumed microblog content in the two networks, including the observation that Chinese respondents considered information from microblogs to be more credible than U.S. respondents. However, the researchers explained “it wasn’t clear whether this difference was due to inherent differences in credibility quality of microblog information in the two cultural contexts, or to the different way that people perceive the credibility of microblog information and/or other information sources.” (p. 581)  

In another study of 746 adults from 76 countries, \textcite{mohd2016correlation} used chi-square analyses to correlate tweet credibility evaluations and demographic attributes (gender, age, education, and location) partitioned into three different settings. That is, location data were partitioned into original setting (Asia, Europe, South America, North America, and Africa), binary setting (Eastern hemisphere, Western hemisphere) and categorical setting (Asia-Pacific, Americas, Europe, Africa). They found a combination of readers' education background and their geo-location had a significant correlation with credibility judgments, but gender and age did not. Further, only location was significantly correlated at all levels of data partitioning. Finally, \textcite{shariff2020review} conducted a literature review of credibility perceptions of online information and concluded that most studies have not paid sufficient attention to how and why readers carry out their credibility judgements of social media posts. Shariff called for more research in credibility perceptions that attends to demographic attributes and other participant characteristics (e.g., membership in particular groups or social media platforms).

Considering all of these findings, in this study, we focused on credibility judgements of health-related social media posts to determine to what extent adult readers relied on their prior beliefs (i.e., knowledge-based validation), evidence types presented in the posts (i.e., discourse-based evaluation) and source characteristics (expertise, gender, and ethnicity) after controlling for the topic of health-related posts.  We also examined whether crowdsourcing platform membership played a role in readers’ credibility evaluations given that the nationalities of members across the two platforms did not overlap.

\section{Methodology}\label{sec:methodology}

\begin{table*}[!htb]
\caption{Independent and control variables in a set of the evaluation tasks for accurate and inaccurate posts.}
\label{tab:var}
\begin{tabular}{|p{8em}|p{15em}|p{17em}|}
\hline
Variable & Description & Manipulation \\
\hline
Independent Variables: & & \\
  \hline
 Evidence Type\textsuperscript{1} & Evidence type supporting the main & A) Research (only for accurate claims)\\
 & claim of the social media post. & B) Testimony  \\
 & & C) Personal experience  \\
 & & D) Consensus  \\
 \hline
 Prior Belief Consistency\textsuperscript{2} & Consistency of self-rated prior beliefs about the topic in relation to the post with the accurate or inaccurate knowledge claim. &  \\
 \hline
Source\textsuperscript{3} & The author of the social media post. & A) Professor of medicine
 \\
 &  & B) Nurse education practitioner
  \\
 &  & C) Professional lifestyle blogger \\
 &  & D) Parent \\
 \hline
Source's Gender\textsuperscript{3} & Self-identified gender of the author  & A) Female \\
& presented in the profile picture.& B) Male \\
\hline
Source's Ethnicity\textsuperscript{3} & Self-identified ethnicity of the author  & A) Asian \\
& presented in the profile picture. & B) Black \\
& & C) White \\
 \hline
 Platform\textsuperscript{4} & Crowdsourcing platform on which & A) Prolific \\
   & participants completed the set of credibility evaluation tasks. & B) Toloka  \\
   \hline
  Control Variable: & & \\
   \hline
   Topic & The topic of the health-related social & A) Fish oil \\
 & media post. & B) Food healthiness  \\
 & & C) Processed red meat  \\
  & & D) Vaccine safety  \\
  & & E) Vitamin D  \\
  \hline
\multicolumn{3}{l}{\multirow{4}{*}{}} \\
\multicolumn{3}{l}{\textsuperscript{1}First-hand evaluation: Discourse-based validation, \textsuperscript{2}First-hand evaluation: Knowledge-based validation,}                 \\
\multicolumn{3}{l}{\textsuperscript{3}Second-hand evaluation,    \textsuperscript{4}Contextual attributes.}              \\   
\end{tabular}
\end{table*}

The present study was intended to understand how adults evaluate short social media posts on health issues. We created accurate (i.e., in line with current scientific knowledge) and inaccurate (i.e., not in line with current scientific knowledge) social media posts on five health topics. Then, we used a computerized algorithm that systematically manipulated evidence type in the posts and source characteristics (the author’s expertise, gender, and ethnicity) to generate 840 unique combinations of the posts. Each unique post was copied 10 times to generate a pool of 8400 posts. The participants, recruited from two crowdsourcing platforms, were asked to evaluate the credibility of ten different posts selected from the larger sample and delivered to each participant via the platform. We examined how evidence type, participants’ prior beliefs, source characteristics of the post and crowdsourcing platform membership were associated with participants' credibility judgments on the posts after controlling for the topic of the posts.

Based on theoretical assumptions supported by previous empirical evidence \autocites{barzilai2020dealing, richter2017comprehension, stadtler2014content}, we assumed that the quality of evidence, participants’ prior beliefs, and expertise of the source would be associated with participants' credibility judgments.  First, we expected that the more that participants’ prior beliefs were consistent with the content in the post, the higher the participants would evaluate the credibility of the post \autocites{mccrudden2016differences, wertgen2023source}.  Second, we expected that posts with claims supported by research-based evidence would be evaluated as more credible than posts that contained evidence from testimony, personal experience, or consensus  \autocites{HornikxJ.M.A2005Aroe,list2021examining}. Third, we expected that posts by sources with relevant domain expertise would be evaluated more credible than posts by sources without domain expertise \autocites{braaten2018task, LIN2016264}. In addition, we were interested in testing the effect of the source’s gender and ethnicity on credibility judgements as well as exploring whether membership in a certain crowdsourcing platform would be associated with participants' credibility judgments. In our analyses, we controlled for the topic of the posts because topic has been shown to play a role in the credibility evaluation of texts  \autocites{braaten2018task, hamalainen2021students}. 

To study these associations, we ran an online quasi-experiment with an incomplete repeated measures design \autocite{shaughnessy2012repeated}. That is, there was no control group, and a single rater rated a maximum of 10 posts out of the 840 unique generated posts. The factors were controlled independently, with practice effects controlled with a completely random order in the Toloka platform, and with balancing in the Prolific platform. The final dataset (8380 posts) consisted of 10 ratings of the 820 unique posts from different raters, and 9 ratings for 20 of the unique posts. All the variables included in this study are presented in Table \ref{tab:var}. 

\subsection{Participants}
We used crowdsourcing to recruit participants for our study. The popularity of crowdsourcing has increased steadily, and it has been used in a variety of fields \autocite{hossain2015crowdsourcing}. A total of 844 participants were recruited from two crowdsourcing platforms: Prolific and Toloka. According to \textcite{chapkovski2023conducting}, Toloka provides access to some populations that can be hard to access through Prolific or MTurk. While the MTurk population is mostly based in the US and India \autocite{difallah2018demographics}, and Prolific users reside mostly in the UK and US \autocite{peer2017beyond}, Toloka’s users mostly reside in Russia and other former Soviet countries. Thus, we used the Prolific and Toloka platforms to obtain a geographically and culturally broad sample. 

The Prolific platform is available for participants who are at least 18 years old from most OECD countries, except for Turkey, Lithuania, Colombia and Costa Rica where Prolific is not available (\autocite{profilic2023participants}). Prolific is also available to participants in South Africa and other participants who are not from OECD supported countries can still participate if they live in a supported country. In September 2023, according to Prolific's audience checker tool \autocite{profilic2023audience}, approximately 58\% of its users identified as female, while 42\% identified as male. Regarding age, 47\% reported being under 30 years old, 48\% were between the ages of 30 and 59, and 5\% were 60 or older. Prolific crowdworkers are primarily recruited by word-of-mouth through social media (e.g., Facebook, Twitter, Reddit) and flyers distributed on university campuses. All participants are paid rewards when their submissions are approved. 

According to its website, the Toloka platform collects data from crowdworkers who live in more than 100 different countries, with a “large concentration of participants coming from Pakistan, Kenya, Brazil, Turkey, India, Egypt, the Philippines, and the United States,” and “about half of Tolokers are native speakers of English, Urdu, Arabic, Russian, or Spanish” \autocite{toloka2023global}. The majority of Toloka workers are between 20–30 years of age; they are equally divided between members of the middle class and working class, and 62\% of its members identify as male. Toloka acquires its users at least through social media presence and advertising. 

Participants were required to complete a demographic survey at the beginning of the study. We used Pew Research's Demographics Questionnaire \autocite{pewresearch2015demographic}, as the basis for our survey. As shown in Table \ref{TableParticipants}, the mean age of respondents was higher in the Prolific platform (\textit{M }= 43) compared to Toloka (\textit{M} = 33) and there were more male respondents in Toloka (56\%) than in Prolific (49\%). In terms of ethnicity, a large percentage of respondents in both platforms identified as White (73\% in Toloka and 85\% in Prolific), fewer identified as Asian (11\% in Toloka and 5\% in Prolific) or as individuals representing multiple ethnicities (10\% in Toloka and 7\% in Prolific) while the smallest number of respondents in either platform identified as Black (6\% in Toloka and 2\% in Prolific). Education levels were similar across the two platforms, but notably, there was no apparent overlap in the nationalities of respondents across the two platforms. In Toloka, more than half (56\%) of respondents identified as Russian, while the remaining respondents lived in regions in Eastern Europe, the Middle East, Kenya, or India. In contrast, more than three-quarters (77\%) of respondents in Prolific were from either the United Kingdom (65\%) or the United States (12\%), and the remaining respondents were from either Canada or other countries in Western and Southern Europe.

\subsection{Generation of  Credibility Evaluation Items for Social Media Posts}\label{sec:evaltask}
\subsubsection{Designing Posts}
We designed short social media posts regarding five health topics on which contradictory information, including accurate and inaccurate claims, had been spread on the internet. In designing the social media posts, we consulted a current Finnish book written by \textcite{knuuti2020kauppatavarana}, a professor and director of the Turku PET Centre, University of Turku. For this book, Knuuti collected 250 common health-related knowledge claims spreading on the internet and analyzed them in light of current scientific knowledge. The chosen topics were the safety of vaccines, Vitamin D, processed red meat, fish oil, and food healthiness. There are several false claims about vaccines on the internet, one of which is that the aluminum in vaccines is dangerous. Also, excessive claims about the benefits of Vitamin D, such as preventing cancer, have been presented. Furthermore, research on the health effects of red processed meat has been misinterpreted by neglecting the potential harms. Finally, there are contradictory claims about the benefits of fish oil and food healthiness. We used research presented in the book to design the research-based social media posts. The sources provided in the book are presented in Table \ref{tab:knuutireferences} of Appendix. 

\begin{figure*}[ht]
\centering
\caption{Example of the credibility evaluation task item on the Prolific platform.}
\includegraphics[width=.60\textwidth]{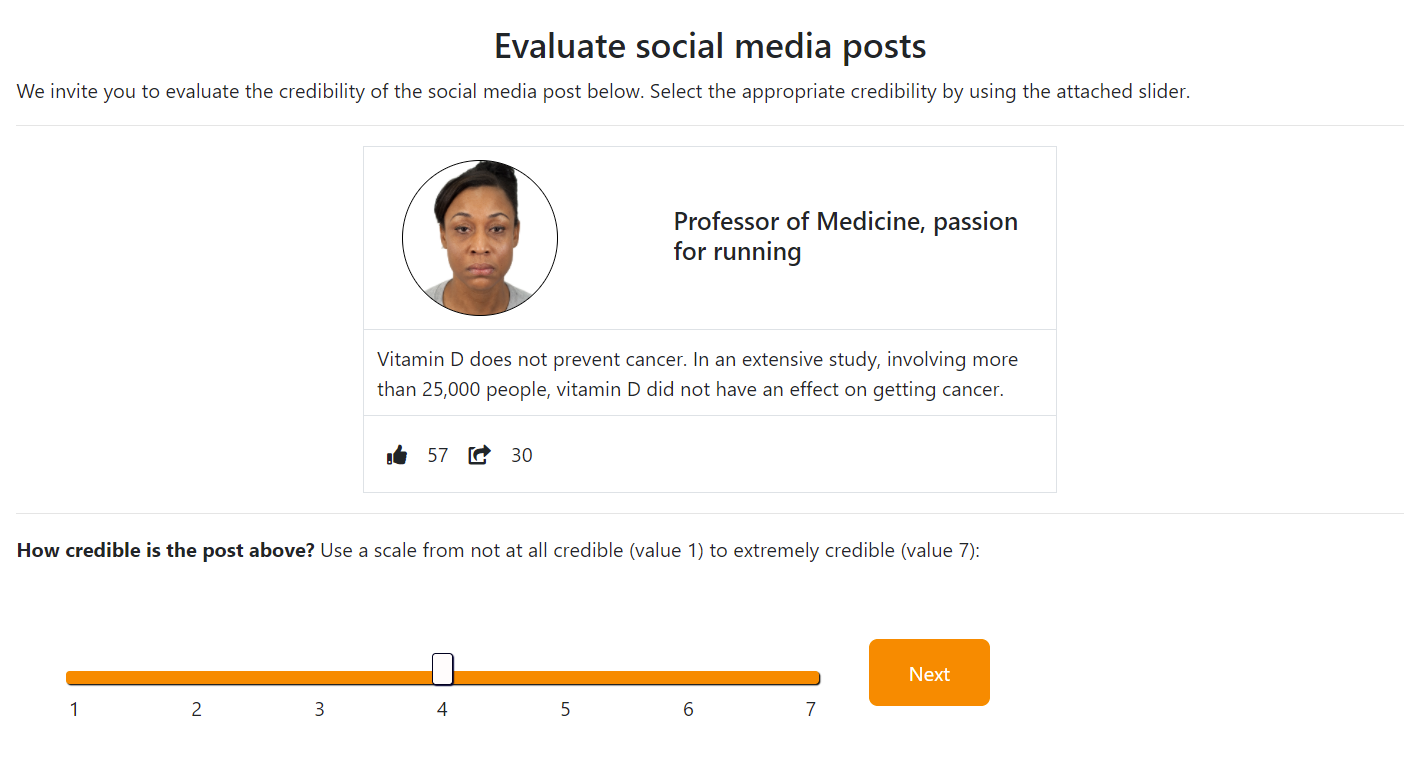}
\label{fig:profilictask}
\end{figure*}

Each post comprised three main components which were systematically manipulated with a computer algorithm: the knowledge claim, supporting evidence, and author of the post (see Figure \ref{fig:profilictask}). Knowledge claims were either accurate, that is, in correspondence with the current scientific knowledge (e.g., Vitamin D does not prevent cancer), or inaccurate (e.g., Vitamin D prevents cancer). The knowledge claims were paired with supporting evidence that represented one of the following evidence types: research evidence, testimony, personal experience, and social consensus \autocites{jacobsen2018thinking, zarefsky2019practice}. Because we used authentic findings as research evidence, we only presented these findings with accurate claims. Examples of the knowledge claims and evidence types related to processed red meat are presented in Table \ref{tab:claimmeat}. 

\begin{table*}[!htb]
\caption{Participants self-reported demographic information by crowdsourcing platform.}.  
\label{TableParticipants}
\begin{tabular}{l|lll}
\hline
     & Toloka & Prolific & Total \\ \hline
Respondents & 425  & 419 & 844\\
Mean age & 33  & 43 & 38 \\
Minimum age &  18 & 20 & \\
Maximum age &  85 & 84  &\\
25 or under & 107 (25\%) & 25 (6\%)& 132 (16\%)\\
26-35 & 139 (33\%) & 127 (30\%) & 266 (32\%)\\
36-45 & 108 (25\%) & 100 (24\%)& 208 (25\%)\\
46-55 & 36 (8\%)& 83 (20\%)& 119 (14\%)\\
over 55 & 35 (8\%) & 84 (20\%)& 119 (14\%)\\
\hline
\textbf{Gender:} &   &  \\
Male &  240 (56\%) & 204 (49\%) &  444 (53\%)\\
Female &  180 (42\%) & 215 (51\%) &  395 (47\%)\\
Other &  5 (1\%) & 0 &  5 (<1\%) \\
\hline
\textbf{Ethnicity:} &   &  \\
Asian & 47 (11\%) & 23 (5\%)  &  70 (8\%)\\
Black & 26 (6\%) &  10 (2\%) &  36 (4\%)\\
White & 309 (73\%) &  357 (85\%) &  666 (79\%)\\
Other or multiple & 43 (10\%) &  29 (7\%) &  72 (9\%)\\
\hline
\textbf{Nationalities}: & Russia – 247 (56\%) & UK – 272 (65\%) & UK – 273 (32\%) \\
& Ukraine – 27 (7\%) & United States – 49 (12\%) & Russia – 247 (29\%) \\
& Turkey – 21 (5\%) &  Italy – 15 (4\%) & United States – 56 (7\%) \\
& Belarus - 20 (5\%)& Spain - 11 (3\%) & Ukraine - 27 (3\%) \\
& Pakistan - 15 (4\%)& Poland - 11 (3\%) & Turkey - 23 (3\%)\\
& India - 15 (4\%)& Canada - 11 (3\%)& Belarus - 20 (2\%)\\
& Kenya - 13 (3\%) & Portugal - 9 (2\%) & India - 15 (2\%) \\
& Other - 67 (16\%) & Other - 41 (10\%) & Other - 183 (22\%)\\
\hline
\textbf{Education:} &   &  \\
Doctorate & 5 (1\%)  & 11 (3\%) & 16 (1\%)\\
Master's degree & 73 (17\%)  & 68 (16\%) & 141 (17\%)\\
Bachelor's degree &  148 (35\%) & 173 (41\%) & 321 (38\%) \\
High school diploma &  107 (25\%) & 123 (29\%) & 230 (27\%)\\
Professional degree & 59 (14\%)  & 17 (4\%) & 76 (9\%)\\
Other qualifications or none & 33 (8\%) & 27 (6\%) & 60 (7\%)\\
\hline
\end{tabular}
\end{table*}

\begin{table*}[ht]
\caption{Arguments of the evaluated social media posts concerning processed red meat.}
\label{tab:claimmeat}
\begin{tabular}{|p{25em}|p{10em}|}
  \hline
Argument with the accurate claim  & Evidence type  \\
  \hline
Red processed meat is a major health hazard. The study review found that red processed meat was associated with a higher incidence and mortality of arterial disease.& Research \\
\hdashline
Red processed meat is a major health hazard. The WHO places processed meat into the highest risk category of carcinogens..& Testimony \\
\hdashline
Red processed meat is a major health hazard. Humans with a balanced diet without red processed meat feel better.& Consensus  \\
\hdashline
Red processed meat is a major health hazard. I eat red meat regularly and at my physical, some of my important indicators are alarming. & Personal Experience \\
   \hline
Argument with the inaccurate claim  &   \\
  \hline
Red processed meat is not a significant health hazard. My personal trainer appreciates red processed meat as a part of a balanced diet. & Testimony \\
\hdashline
Red processed meat is not a significant health hazard. Humans are carnivores and this is how we have survived. & Consensus  \\
\hdashline
Red processed meat is not a significant health hazard. I eat red meat regularly and at my physical, all of my important indicators are great. & Personal Experience \\
\hline
\end{tabular}
\end{table*}

\begin{figure*}[ht]
\centering
\caption{Illustration of algorithmic process\textsuperscript{1} used to generate 840 unique combinations of social media posts.}
\includegraphics[width=.99\textwidth]{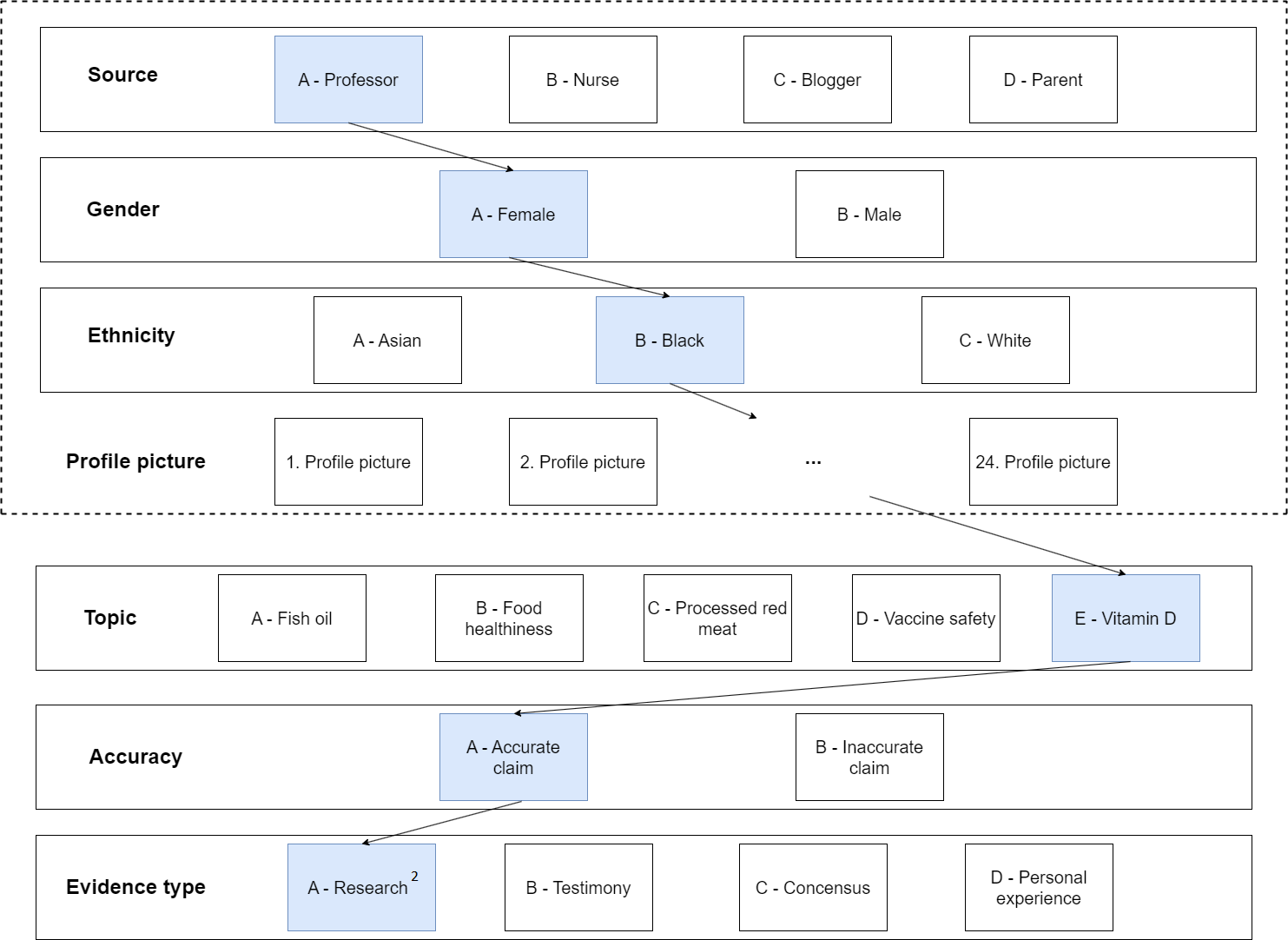}
    \bigskip
    \begin{minipage}{\textwidth}
        \captionsetup{labelformat=empty}
        \raisebox{1.5ex}{\llap{\textbf{}}}:\newline
        Note: \textsuperscript{1} Rows 1-3 represent steps in the computerized process of selecting a unique combination of source features (source, gender, and ethnicity [i.e., Professor who is female and Black] paired with a profile picture matching those features. Rows 4-6 represent steps in generating content of the social media post associated with that source by selecting a topic (Vitamin D), claim type (accurate), and evidence type (research).\newline
        \textsuperscript{2} The Research type of evidence was only available for accurate claims.
    \end{minipage}
\label{fig:generation}
\end{figure*}

We created four types of authors (or sources) and varied their expertise. Two sources, a professor of medicine and a nurse education practitioner for undergraduate students, were experts in the health domain. The other two sources, a professional lifestyle blogger and a parent, did not have domain expertise. In addition to this information, the author’s profile included a short description, such as ‘passion for running’, to make the social media messages seem more authentic. 

Because some prior work has suggested that an author's ethnic background and gender may be associated with readers' responses to social media health information \autocites{armstrong2009blogs, spence2013intercultural}, we also varied the author gender and ethnicity. To do this, we used the Chicago face database \autocite{ma2015chicago}, which provides facial pictures along with the self-identified ethnicity and gender, as well as a subjective attractiveness score based on ratings from US-based raters. Using this data, we selected pictures of people with average attractiveness who were between 35 and 45 years of age. We included males and females, as well as those who self-identified as Asian, Black, and White. The age and attractiveness selection were performed to create a uniform set of facial pictures and thus rule out very unlikely combinations, such as 20 year old professors. The selected profile pictures are shown in Appendix (Figure \ref{fig:faces}).

\subsubsection{Full Factorial Design}

Next, unique combinations of sourcing features (author expertise, gender, ethnicity, and profile picture) and content features (topic, claim type, and evidence type) were generated by applying mathematical principles of combinatorics \autocite{kuhn2009combinatorial} to a full factorial design to study all possible combinations of the different levels of factors being considered. Traditional experimental designs in credibility evaluation often avoid full factorial design because the total number of all combinations can escalate quickly, surpassing what is feasible to study - hence, the complexities we touch upon in the paper title. For instance, an experiment may examine two factors: source expertise of a post and evidence backing the post. If each factor has two levels (e.g., source expertise includes 1) Professor and 2) Blogger, and evidence type comprises 1) Research and 2) Personal experience), this results in four unique combinations (2x2). If there are three levels for each factor, then nine (3x3) combinations arise. Introducing a third factor with three potential values, such as source ethnicity, balloons the number of combinations to 27 (3x3x3). The number of factors and levels renders full factorial design  seemingly unfeasible in studies of credibility evaluation. 

In our study, we implemented a full factorial design that tested 840 unique combinations. With such a vast number of combinations, we encountered two main challenges. The first was the effort of creating 840 distinct social media posts and ensuring that each combination was created just once. The second challenge was how to conduct an experiment with this multitude of posts. We addressed the first issue using an algorithm and software that automatically generated the required 840 social media posts, ensuring all combinations were produced. This resulted in 480 unique posts with accurate knowledge claims and 360 unique posts with inaccurate knowledge claims. For the second challenge, we employed two crowdsourcing platforms, from which large numbers of participants could easily be recruited.  To conduct the experiment, we used an incomplete repeated measures design since expecting one person to evaluate all 840 posts would be unrealistic. As a result, each participant only assessed 10 unique posts.

Figure \ref{fig:generation} illustrates how social media posts were generated with this computerized procedure by using the credibility evaluation task item presented in Figure 1 as an example. A combination of ethnicity, gender, and profession was used to select the profile picture (i.e., Black female professor) associated with the social media post, which was generated using a combination of the topic (Vitamin D), accuracy of claim (accurate), and evidence type (research). 

Although we conceptualized perceived credibility as a multidimensional construct, reflecting the idea of first- and second hand evaluation \autocites{barzilai2020dealing, stadtler2014content}, we measured participants’ credibility judgments with a single item  (cf. \cite{hanimann2023believing}). Participants were assigned to evaluate the credibility of ten posts on a single item scale from 1 to 7 (1 = not at all credible; 7 = extremely credible). The single item was used because we sought to understand how different content- and source-related issues affected participants’ overall evaluations of the social media posts’ credibility. Single-item scale may be advantageous because participants may resent being asked questions that appear to be repetitive \autocite{fuchs2009using}. In this study, the unit of analysis was the credibility evaluation of one post.

\subsection{Prior Belief Measure}
Before gaining access to the credibility evaluation task, participants’ prior beliefs on the topic were measured using one statement for each topic at hand and a 7-point scale. The statements were worded, for example, ‘Vaccines are safe', and `Red meat is healthy'. We coded the responses to prior belief items in relation to how consistent they were with the claim. For accurate claims, a prior belief score of 7 corresponded to a prior belief consistency score of 3, and a prior belief score of 1 corresponded to a    consistency score of -3. For inaccurate claims, this was done in reverse, with a prior belief score of 1 corresponding to consistency score of 3 and a prior belief score of 7 corresponding to a consistency score of -3. For example, if the participant responded with 1 on the item “Vaccines are safe,”, it was coded as -3 for the accurate claim. In the case of an inaccurate claim, the participant’s response 1 was coded as 3. The variable was labeled as prior belief consistency, with a scale from -3 to 3.

\subsection{Data Collection Procedure}
Prior work has established cultural differences in credibility evaluation of online reviews \autocite{venturebeat2022how}, as well as how people use and trust health information in online contexts \autocites{khosrowjerdi2020national, song2016trusting}. Because we opted for two crowdsourcing platforms, with a wide variety of nationalities in the userbase, crowdworkers on Toloka and Prolific were offered an opportunity to take part in our study. Participants were first asked to fill out a questionnaire on demographic information and the prior belief measure. After completing the questionnaire successfully, the participants proceeded to the credibility evaluation task, in which they were asked to evaluate the credibility of ten social media posts. We assigned only ten credibility evaluation items to each participant to ensure that they could focus sufficiently on each item. In addition, with this procedure, we could recruit a considerable number of participants which assured that they represented diverse backgrounds. 

On the two platforms, somewhat different methodologies were used to deliver the set of credibility evaluation items to raters because of how these platforms worked. On the Prolific platform, we gave the tasks to participants in predetermined batches of ten, with minimized repetition of the manipulated aspects. In principle, ten posts could have some repetitive elements. The same topic, evidence type, source, and profile picture could have been shown to the participant in a maximum of three posts. However, the exact same post was not shown twice to the same participant. On the Toloka platform, each participant had an opportunity to make ten evaluations that were assigned to them at random. This was done by randomizing the order of posts that were given to the crowdworkers on a first-come first-serve basis. However, like Prolific, the exact same post was not shown twice to the same participant. Thus, the items were manipulated independently of each other. Only the top 20 \% performing crowdworkers on the Toloka platform were offered the chance to participate in the study. Crowdworkers’ performance was calculated by their success in completing tasks on the platform (i.e., the acceptance rate of their previous work at the platform). All evaluations made on the Toloka platform were accepted. Of the Toloka participants, $ 99 \% $ completed the maximum of ten evaluations. The task used on the Prolific platform is shown in Figure \ref{fig:profilictask}.

\subsection{Statistical Analysis}\label{sec:statanalysis}

In the statistical analysis, the unit of analysis was the credibility evaluation of one social media post. Analyses were conducted separately for posts including accurate and inaccurate knowledge claims, which are later referred to as accurate and inaccurate posts below. Data consisted of 4787 credibility evaluations of accurate posts and 3593 evaluations of inaccurate posts. Twenty credibility evaluations were excluded from the analysis due to technical problems.  

All statistical analyses were conducted using R. To examine how independent variables (see Table \ref{tab:var}) were associated with the credibility evaluations of social media posts, we used a cumulative link mixed model. The cumulative link mixed model (CLMM), also known as ordered regression model \autocite{christensen2015ordinal}, has been advocated for use with ordered outcome measures \autocite{fullerton2021ordered}. 

To provide a CLMM model and odds ratios, we used the ordinal package created by \textcite{christensen2015ordinal} for the R programming language. The specific formula used is follows: "Credibility\textasciitilde  Accuracy + ClaimEvidence + Topic + Author + Platform + Gender + Ethnicity + (1|ID)". The control variable ID corresponds to each individual evaluator who performed a maximum of ten evaluations in the sample. For the reference category, that is, the category other categories are compared to in the model (see reference categories in Tables \ref{TableAcc} and \ref{TableInacc}), we chose the category that one would assume to produce the highest credibility rating. For all other variables, we used the category that was first in alphabetical order. Thus, for evidence type, we chose research; for prior beliefs, we chose highest consistency, and for source expertise, we chose professor. For gender, ethnicity, platform, and topic, we used alphabetic order.

We also present odds ratios (ORs), which are calculated with the ordinal package \autocite{christensen2015ordinal}, to highlight the association between the independent variable and the dependent variable. Odds ratios have been advocated for use as an effect size measure with cumulative link mixed models and ordered regression models \autocites{fleiss1994measures, fullerton2016ordered}. Moreover, odds ratios have been used with CLMM models in a variety of fields \autocites{oehm2022identifying, rezapour2021modeling}. An OR greater than 1 indicates how many times greater the expected credibility evaluation value of the category of interest is compared to the reference category. In contrast, an OR smaller than 1 indicates that the expected evaluation value in the reference category is greater than that in the category of interest. The ORs were transformed to another popular effect size measure, Cohen's d, with the effectsize R package created by \textcite{ben2020effectsize}. The Cohen's d effect sizes were interpreted as follows: small (\textit{d} = 0.2), medium (\textit{d} = 0.5), and large (\textit{d} = 0.8)\autocite{cohen2013statistical}. To calculate the goodness-of-fit statistics, we used the anova.clmm function of the RVAideMemoire package \autocite{herve2020package} to perform a likelihood-ratio test \autocite{crainiceanu2004likelihood}.

\section{Results}\label{sec:results}

\begin{table*}[!htb]
\centering
\caption{Descriptives for the credibility evaluations by independent and control variables.}
\label{TableDesc}
\begin{tabular}{|l|p{9em}|p{9em}|p{6em}|p{6em}|p{6em}|}
\hline
Variable & Manipulation &Mean\textsuperscript{1}  & SD & N (8380)\textsuperscript{2} \\  \hline
 Accuracy of the Claim
 & Accurate & 4.25 & 1.72 & 4787 \\
 & Inaccurate & 3.65 & 1.81 & 3593 \\
\hdashline
Evidence Type\textsuperscript{3} 
 & Research & 4.36 & 1.64 & 1193\\
 & Testimony  & 4.13 & 1.77 & 2400\\
 & Consensus & 3.94 & 1.81 & 2389\\
& Personal experience & 3.74 & 1.80 & 2398\\
\hdashline
Prior Belief Consistency\textsuperscript{4}
 & 3 & 4.30 & 1.90 & 1120 \\
 & 2 & 4.18 & 1.71 & 1162 \\
 & 1 & 4.16 & 1.65 & 1276 \\
 & 0 & 4.02 & 1.69 & 1583 \\
 & -1 & 3.82 & 1.73 & 1198 \\
 & -2 & 3.68 & 1.78 & 1083 \\
 & -3 & 3.75 & 2.02 & 958 \\
\hdashline
Source\textsuperscript{5} & Professor of medicine   & 4.40 & 1.77 & 2094\\
& Nurse education practitioner  & 4.22 & 1.69 & 2095\\
& Professional lifestyle blogger  & 3.80 & 1.78 & 2093\\
& Parent  & 3.56 & 1.78 & 2098\\
\hdashline
Source's Gender\textsuperscript{5} 
 & Female & 4.02 & 1.80 & 4191\\
 & Male & 3.97 & 1.77 & 4189\\
\hdashline
Source's Ethnicity\textsuperscript{5}
& Asian & 3.97 & 1.79    & 2797 \\
& Black & 4.05 & 1.77 & 2789\\
& White & 3.97 & 1.79 & 2794 \\
\hdashline
Platform\textsuperscript{6} 
 & Prolific & 3.68 & 1.75 & 4190 \\
 & Toloka & 4.32 & 1.77 & 4190 \\
\hline
Topic
& Food healthiness & 4.22 & 1.76 & 1673 \\
& Vitamin D & 4.04 & 1.84 & 1680 \\
& Fish oil & 4.01 & 1.68 & 1676 \\
& Processed red meat & 3.99 & 1.74 & 1678\\
& Vaccine safety & 3.72 & 1.87 &1673 \\
 \hline
  \hline
\multicolumn{5}{l}{\multirow{4}{*}{}} \\
\multicolumn{5}{l}{\textit{Note.} \textsuperscript{1}Credibility evaluations were given on a scale from 1 to 7. \textsuperscript{2}N is the number of}\\
\multicolumn{5}{l}{credibility evaluations within each variable.}\\
\multicolumn{5}{l}{\textsuperscript{3}First-hand evaluation: Discourse-based validation, \textsuperscript{4}First-hand evaluation: Knowledge-based validation,}                 \\
\multicolumn{5}{l}{\textsuperscript{5}Second-hand evaluation,    \textsuperscript{6}Contextual attributes.}              \\   
\end{tabular}
\end{table*}
 

\begin{table*}[!htb]
\centering
\caption{CLMM results for accurate social media posts.}
\label{TableAcc}
\begin{tabular}{|p{19em}|p{5em}|p{5em}|p{5em}|p{5em}|}
\hline
\textbf{Fixed effect terms} & \textbf{Estimate} & \textbf{Std. error}   & \textbf{Odds Ratio} & \textbf{Cohen's d} \\ 
\hline
\textbf{Independent variables:}  & & & & \\ 
Evidence Type\textsuperscript{1} – Reference: Research & & & & \\
Testimony    &                                               0.19*  & 0.08  &  1.21 & 0.11\\
Consensus               &       -0.29***  & 0.08 & 0.75 & -0.16\\
Personal experience             &               -0.33***  &  0.10 & 0.72 & -0.18\\
\hdashline
Prior Belief Consistency\textsuperscript{2} – Reference: 3 & & & & \\
 2            &          -0.45***  &  0.12  & 0.64 & -0.25 \\
 1                    &      -0.68***  &  0.12 & 0.50 & -0.37\\
 0     &      -1.19***  &  0.11   & 0.30 & -0.65\\
 -1   &    -1.21***  &  0.13  & 0.30 & -0.66\\
 -2  & -1.48***  &  0.15 & 0.23 & -0.82\\
 -3   &     -1.28***  &  0.14   & 0.28 & -0.71\\
\hdashline
Source's Expertise\textsuperscript{3} – Reference: Professor of medicine &   &    &  &  \\
Nurse education practitioner  &  -0.28***   &  0.08  & 0.75 &  -0.16 \\
Professional lifestyle blogger   &  -0.89***   & 0.08   & 0.41 & -0.49 \\
Parent   &     -1.13***   &  0.08  & 0.32 & -0.62 \\
\hdashline
Source's Gender\textsuperscript{3} – Reference: Female & & & & \\
Male                 &        -0.04  & 0.06 & 0.96 & -0.02\\
\hdashline
Source's Ethnicity\textsuperscript{3} – Reference: Asian & & & & \\
Black        &            0.10  &  0.07  & 1.11 & 0.06\\ 
White      &               0.08 &  0.07   &  1.08& 0.04\\ 
\hdashline
Platform\textsuperscript{4} - Reference: Prolific &     &    &  &  \\
Toloka                                     &     0.69***   & 0.09    & 2.00 & 0.38\\
\hline
\textbf{Control variable:}  & & & & \\
Topic – Reference: Fish oil & &  & &\\
Food healthiness                                                             &     1.49***  & 0.09 & 4.44 & 0.82\\
Processed red meat                                                            &      0.69***  & 0.09  & 1.99 & 0.38\\
Vaccine safety                                                         &      0.39***  & 0.09  & 1.47 & 0.21\\
Vitamin D                                                        &      1.57***  & 0.11 & 4.80 & 0.86\\
\hline
\textbf{Post-Hoc analysis} – Evidence Type: &  &  & &\\
Concensus – Testimony  &  -0.48*** & 0.08 & 0.62 & -0.26 \\
Concensus – Personal experience  & 0.04 & 0.08 & 1.04 & 0.02 \\
Personal experience – Testimony &  -0.52*** & 0.09 & 0.60 & -0.28\\ 
\hline
 \textbf{Post-Hoc analysis} – Source's expertise: &  &  & &\\
Nurse education practitioner – Professional lifestyle blogger & 0.60*** & 0.08 & 1.82 & 0.33\\
Nurse education practitioner – Parent & 0.85*** & 0.08 & 2.34 & 0.47\\
Parent – Professional lifestyle blogger & -0.25***  & 0.08 & 0.80 & -0.14\\
 \hline
\multicolumn{5}{l}{\multirow{4}{*}{}} \\
\multicolumn{5}{l}\textit{Note.} ** p < 0.01 ***p < 0.001. {\textsuperscript{1}First-hand evaluation: Discourse-based validation, \textsuperscript{2}First-hand evaluation:}                 \\
\multicolumn{5}{l}{Knowledge-based validation, \textsuperscript{3}Second-hand evaluation,    \textsuperscript{4}Contextual attributes.}              \\   
\end{tabular}
\end{table*}

\begin{table*}[!htb]
\centering
\caption{CLMM results for inaccurate social media posts.}
\label{TableInacc}
\begin{tabular}{|p{19em}|p{5em}|p{5em}|p{5em}|p{5em}|}
\hline
\textbf{Fixed effect terms} & \textbf{Estimate} & \textbf{std. error}   & \textbf{Odds Ratio} & \textbf{Cohen's d} \\ 
\hline
\textbf{Independent variables:}  & & & & \\ 
Evidence Type\textsuperscript{1} – Reference: Testimony & & & & \\ 
Consensus & 0.01 & 0.09& 1.01 & 0.01 \\ 
Personal experience & -0.32***& 0.10& 0.72 & -0.18 \\ 
\hdashline
Prior Belief Consistency\textsuperscript{2} – Reference: 3 & & & & \\
 2 & -0.23 & 0.17 & 0.79 & -0.12\\ 
 1  & -0.61*** & 0.17 & 0.54 & -0.34\\ 
 0  & -0.85*** & 0.16 & 0.43 & -0.47\\ 
 -1  & -1.31***& 0.17 & 0.27 & -0.72\\ 
 -2 & -1.54*** & 0.17 & 0.21 & -0.85\\ 
 -3 & -2.12*** & 0.18 & 0.12 & -1.17\\ 
\hdashline
Source's Expertise\textsuperscript{3}  – Reference: Professor of medicine &   &    &  &  \\
Nurse education practitioner  & -0.08 & 0.09 & 0.93 & -0.04\\ 
Professional lifestyle blogger   & -0.71*** & 0.10 & 0.49 & -0.39 \\ 
Parent    & -0.86*** & 0.10 & 0.42 & -0.47\\ 
\hdashline
Source's Gender\textsuperscript{3} – Reference: Female & & & & \\
Male  & -0.10 & 0.07 & 0.91 & -0.05\\ 
\hdashline
Source's Ethnicity\textsuperscript{3} – Reference: Asian & & & & \\
Black  & 0.11 & 0.08 & 1.12& 0.06\\ 
White  & -0.07 & 0.08 & 0.93 & -0.04\\ 
\hdashline
Platform\textsuperscript{4} – Reference: Prolific &     &    &  &  \\
Toloka  & 1.04*** & 0.11 & 2.83 & 0.57\\ 
\hline
\textbf{Control variable:}  & & & & \\
Topic – Reference: Fish oil & &  & &\\
Food healthiness   & -1.32 & 0.11 & 0.27 & -0.73\\ 
Processed red meat  & -0.90& 0.11 & 0.41 & -0.50\\ 
Vaccine safety  & -1.46 & 0.11 & 0.23 & -0.80\\ 
Vitamin D  & -2.21 &0.16 & 0.11 & -1.22\\ 

\hline
\textbf{Post-Hoc analysis} – Evidence Type: &  &  & &\\
Consensus – Observation  & 0.35*** & 0.08 & 1.42 & 0.19 \\
\hline
 \textbf{Post-Hoc analysis} – Source's expertise: &  &  & &\\
Nurse education practitioner – Professional lifestyle blogger & 0.61*** & 0.09 & 1.85 & 0.34\\
Nurse education practitioner – Parent & 0.81*** & 0.09 & 2.24 & 0.44\\
Parent – Professional lifestyle blogger & -0.19 & 0.09 & 0.82 & -0.11\\
 \hline
\multicolumn{5}{l}{\multirow{4}{*}{}} \\
\multicolumn{5}{l}{\textit{Note.} ***p < 0.001.  \textsuperscript{1}First-hand evaluation: Discourse-based validation, \textsuperscript{2}First-hand evaluation: Knowledge-based}   \\
\multicolumn{5}{l}{validation, \textsuperscript{3}Second-hand evaluation,    \textsuperscript{4}Contextual attributes.}              \\   

\end{tabular}
\end{table*}

In this study, we were interested in how content-related issues (i.e., prior belief consistency and evidence type), source characteristics (i.e., source expertise, ethnicity and gender), and crowdsourcing platform (i.e. Toloka or Prolific) were associated with reader's credibility evaluations of short social media posts when the posts' topic were controlled for. Specifically, we were interested in the relative impact of the identified factors on the perceived credibility of social media posts containing accurate or inaccurate health claims. Descriptive statistics are shown in Table \ref{TableDesc}. We ran the CLMM analyses separately for credibility evaluations of the posts with accurate and inaccurate knowledge claims. A Likelihood-ratio test between full models and their null model counterparts, that is, models without predictors, showed a good fit, with both being statistically significant at the \textit{p} < .001 level.

Results of the CLMM analysis are presented in Table \ref{TableAcc} for the posts with accurate claims and in Table \ref{TableInacc} for the posts with inaccurate claims. As assumed, the content-related issues (i.e., prior belief consistency and evidence type) affected readers' credibility evaluations of the posts regardless of whether the claims were accurate or inaccurate. As shown in Tables \ref{TableAcc} and \ref{TableInacc}, participants' prior beliefs significantly affected their credibility evaluations. The more consistent participants' prior beliefs were with the claim of the post, the more the message was evaluated as being credible. Effect sizes (Cohen's d) between evaluations of the very high consistency posts and the very low consistency posts were -0.71 for posts with accurate claims and -1.17 for posts with inaccurate claims. 
 
The role of evidence type in credibility evaluation was small. When other evidence types were compared to research evidence (posts with accurate claims), the effect sizes ranged from 0.11 to -0.18, and when compared to testimony (posts with inaccurate claims), the effect sizes ranged from 0.01 to -0.18. Notably, for the posts with accurate claims, the posts with testimony were evaluated the most credible, testimony being 1.21 times more likely to be evaluated credible than research evidence. When evaluating posts with inaccurate claims, readers evaluated the posts with testimony and consensus similar in credibility.

In terms of the source characteristics, the expertise of the source affected credibility evaluations, whereas source gender and ethnicity did not have any effect. The posts written by the expert sources, that is professor of medicine or nurse education practitioner, were evaluated to be more credible than posts by non-expert sources. Cohen's d varied from -0.43 to -0.62 when the expert sources were compared to the parent and from -0.39 to -0.49 when the expert sources were compared to the professional lifestyle blogger.

In terms of contextual attributes of the crowdworkers themselves, the choice of data collection platform had a notable impact on our results. Participants from the Toloka platform tended to perceive both kinds of posts as more credible, whether they contained accurate or inaccurate claims. 
This disparity between the platforms was even more pronounced for posts containing inaccurate claims (d = 0.57), with a post of this nature being 2.83 times more likely to be rated higher on the Toloka platform compared to Prolific platform. 

\begin{figure*}[!htb]
\centering
\caption{Credibility evaluations for accurate (left) and inaccurate (right) posts by source and prior belief consistency.}
\includegraphics[width=.45\textwidth]{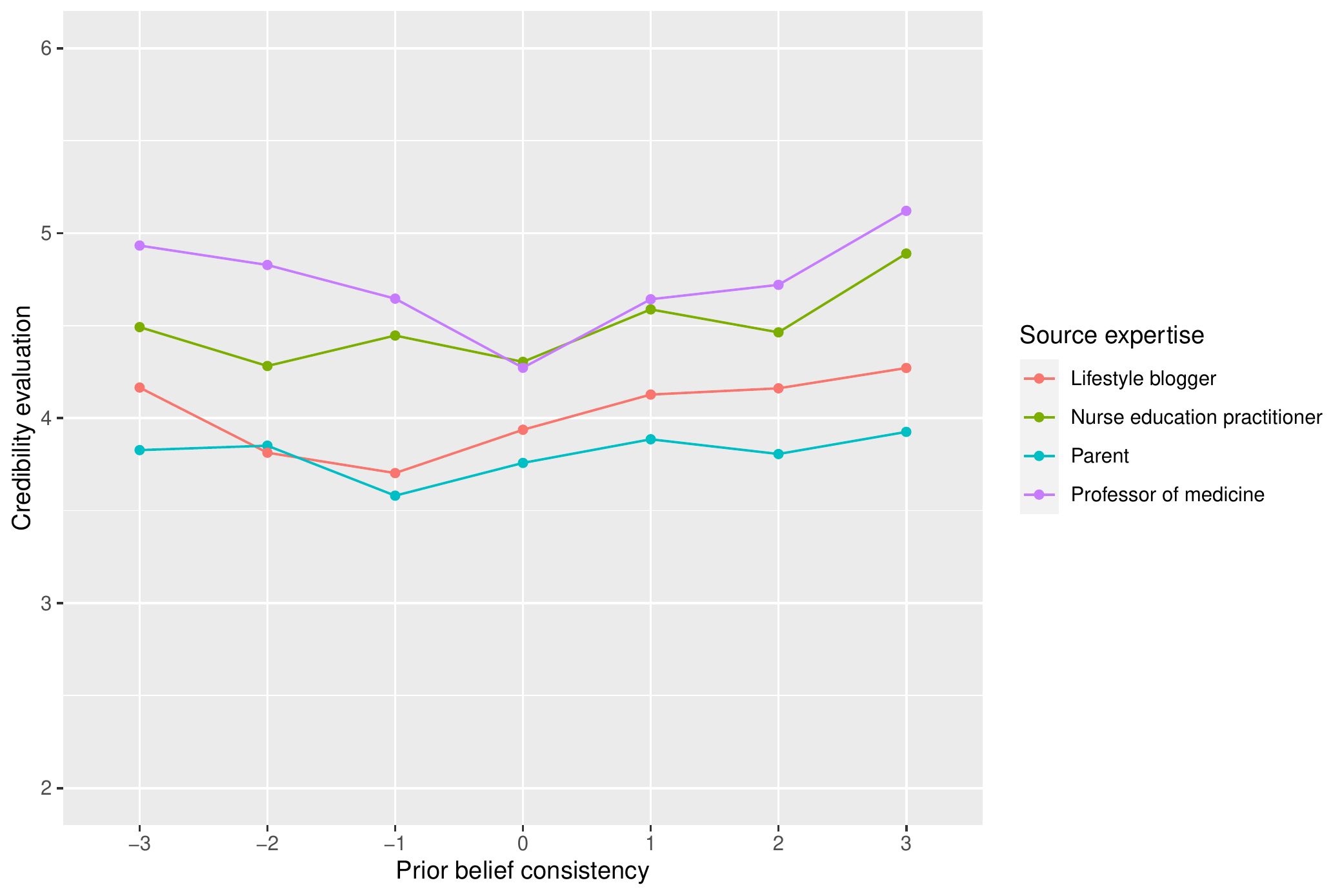}
\includegraphics[width=.45\textwidth]{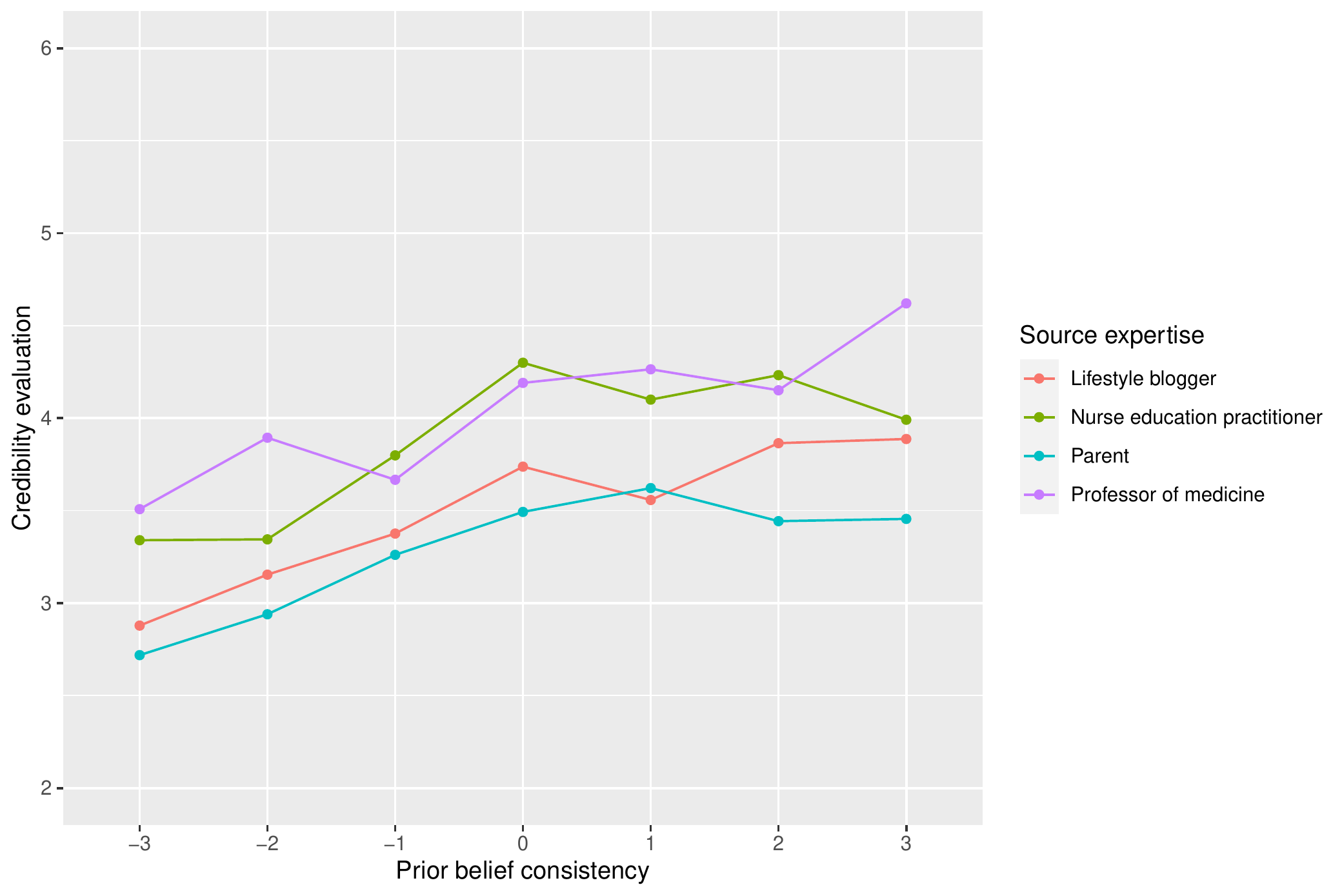}
\label{fig:pbsource}
\end{figure*}

\begin{figure*}[!htb]
\centering
\caption{Credibility evaluations for accurate (left) and inaccurate (right) posts by source and evidence type.}
\centering
\label{fig:evidencesource}
\includegraphics[width=.45\textwidth]{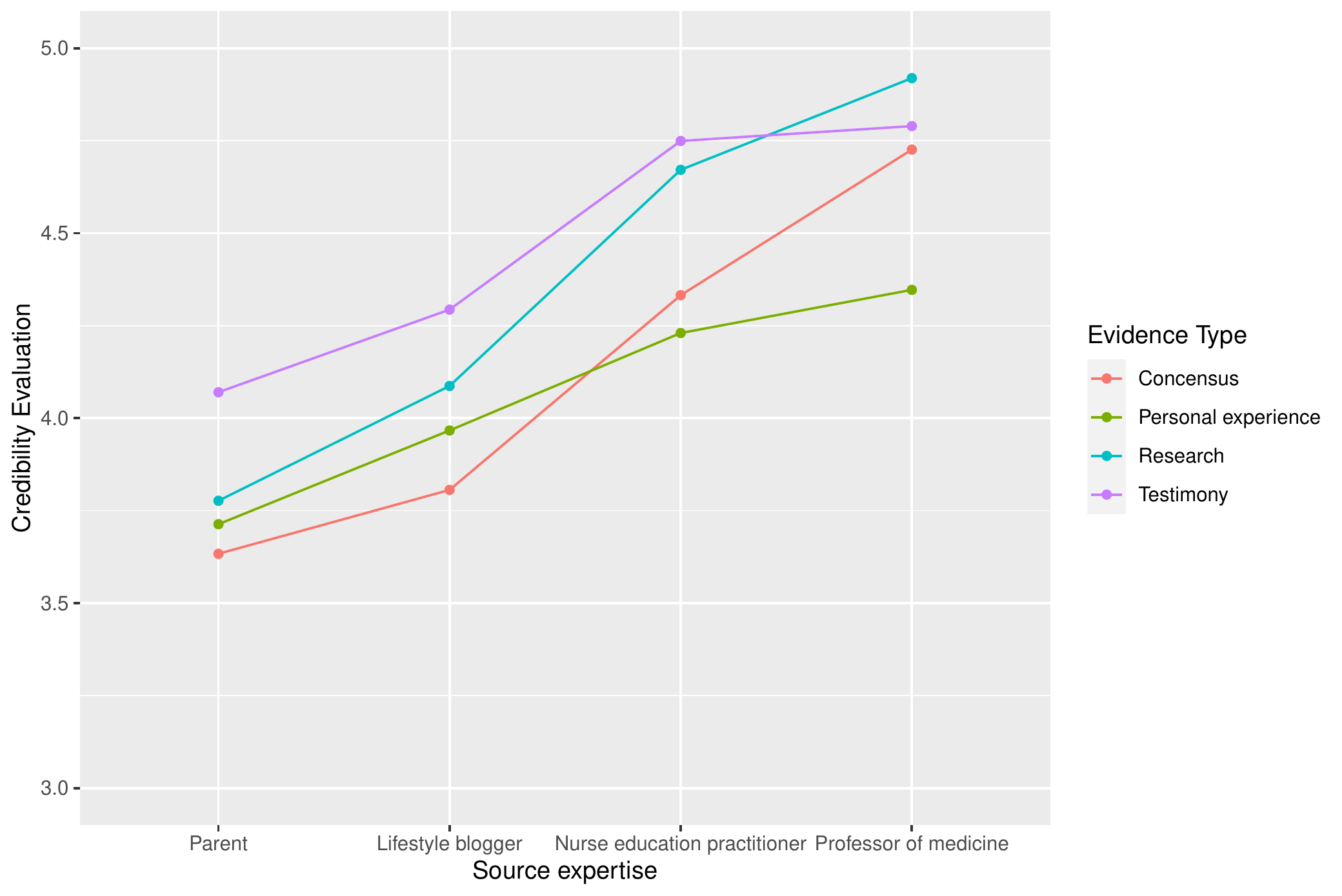}
\includegraphics[width=.45\textwidth]{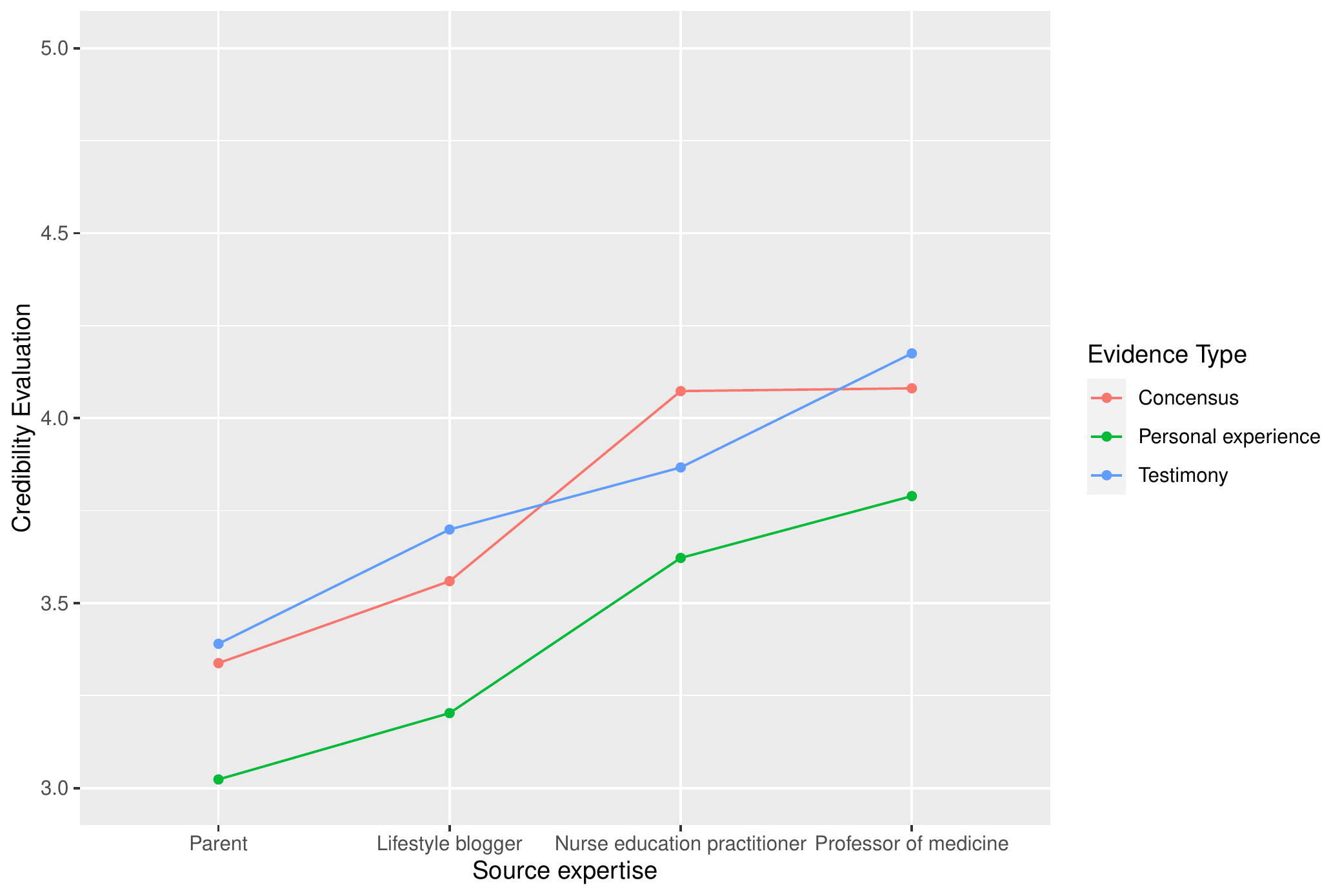}
\label{fig:patterns}
\end{figure*}

To sum up, our results suggest that adults' prior beliefs play a considerable role in the credibility evaluation of short social media posts across different sources (See Figure \ref{fig:pbsource}). This is particularly significant when adults face inaccurate information on social media. Furthermore, adult readers seem to pay more attention to source features indicating the source's expertise than to the quality of the evidence (See Figure \ref{fig:evidencesource}). For example, regardless of evidence type used in the arguments, participants trusted the posts by the professor of medicine more than those by the parent. Finally, adults’ credibility evaluations differed significantly across the two crowdsourcing platforms, such that crowdworkers in the Toloka platform tended to evaluate the credibility of the inaccurate posts, in particular, higher than crowdworkers in the Prolific platform. 

\newpage
\section{Discussion}\label{sec:discussion}

This study sought to understand how adult readers evaluate the credibility of short health-related social media posts that vary in source characteristics and argument quality. Our design of the social media posts included some unique features. First, we chose topics on which contradictory claims, including accurate and inaccurate information, spread on the internet, by relying on the thorough investigation of a medical professor who has collected scientific evidence to combat the common misinformation on health issues \autocite{knuuti2020kauppatavarana}. Second, this study was the first to manipulate evidence type (i.e., research evidence, testimony, consensus, and personal experience) in social media posts showing that the quality of evidence had a minor effect on adult readers’ credibility evaluations. Another unique feature of this study was that we recruited participants from two different crowdsourcing platforms to have a culturally broad sample, which had interesting results. First, we discuss the findings in terms of first-and second-hand evaluation (i.e., content and source evaluation), after which we discuss effects of the crowdsourcing platform as a context for evaluating social media posts. Finally, we discuss how these findings provide important contributions to theoretical models and methodologies for studying credibility evaluation and consider limitations of the study.    

In terms of first-hand evaluation strategies, participants' prior beliefs mattered more regarding their credibility judgments than the quality of the evidence presented in the post. The more consistent the participants' prior beliefs were with the claim of the post, the more credible the post was evaluated as being. This association was found in credibility judgments of both accurate and inaccurate posts, and it is in line with previous research \autocites{abendroth2020mere, mccrudden2016differences}. Relying on one's prior beliefs in credibility evaluation is, however, a double-edged sword. When a person has accurate prior beliefs, they can be effectively used to disregard misinformation. In contrast, when a person has false prior beliefs, relying on them in credibility evaluation is unwise, as this strategy may do nothing but strengthen one's false beliefs. Therefore, it is essential that, in our increasingly complex knowledge society, laypeople acknowledge the limits of their prior knowledge and beliefs when consuming online information \autocite{osborne2022science}. This becomes even more important when comprehending health information in social media contexts, given the level of domain knowledge required to understand complicated medical terminology and evidence that misinformation on major public health issues is very common on social media platforms (\cite{suarez2021prevalence}; see also \cite{vosoughi2018spread}).

Readers' prior beliefs are a quick internal resource that they can easily access and use in their credibility evaluations. In contrast, using discourse-based validation strategies, such as evaluating the strength of the argument, may require more processing effort. However, it is still somewhat surprising that the evidence type had relatively small effect (\textit{d}s 0.02–0.28) on participants' evaluation of accurate social media posts in which all four evidence types (i.e., research, testimony, consensus, and personal experience) were represented. It is understandable that research and testimony had a similar effect on credibility evaluation because the testimonies used in our posts relied on authorities (e.g., European Food Safety) or experts (e.g., a doctor), but it is surprising that the posts relying on consensus or personal experience were valued rather similarly to the posts that included research evidence or expert testimony.  

It is true that findings in previous studies by \textcite{list2021examining}  and \textcite{HornikxJ.M.A2005Aroe} indicate that research-based evidence is highly valued among readers. However, in these studies participants evaluated evidence in more traditional contexts, such as newspaper stories \autocite{list2021examining}, not social media. Our study uniquely examined evidence as part of social media posts, suggesting that research-based evidence may be less valued in the social media context. One reason may be that narrative ways of knowing, such as personal experience, are more persuasive in social media. Instead of seeking research-based evidence for decision-making, people may turn to other consumers’ or influencers’ experiences that they can relate to \autocite{feng2021expert}. Such behavior may also be reflected in the credibility evaluation of the social media posts. Another explanation is that many students leave high school unable to understand, evaluate, or write arguments, and especially scientific arguments \autocites{duschl2002supporting, takao2003assessment, larson2009improving}   Strategies for comprehending and evaluating arguments are not explicitly taught in school and are rarely emphasized in university curricula \autocite{osborne2010arguing} and deficits in aspects of scientific reasoning are pervasive among university students  \autocite{britt2014scientific}. Thus, more research is needed to understand if increased exposure to arguments, and explicit instruction and practice in the skills of argumentation, influence adults’ discourse-based credibility evaluation skills, particularly in social media contexts.  

In terms of second-hand evaluation strategies, we manipulated the source's expertise, gender, and ethnicity. Our results suggest that participants focused primarily on a source's expertise when evaluating the credibility of social media posts. In fact, a source's gender and ethnicity were not associated at all with participants' credibility judgments. Our findings about the role of source expertise in credibility evaluation are in line with previous research (e.g., \textcite{LIN2016264}). Participants evaluated the expert sources (i.e., the professor of medicine and nurse education practitioner) as being more credible than the laypersons (i.e., the lifestyle blogger and parent) almost regardless of the presented evidence (Figure \ref{fig:evidencesource}). This could be explained based on default trust: the default trust occurs when there is no reason to doubt the argument of the source. This kind of trust is typically based on inductive reasoning \autocite{shieber2015testimony}. For example, trust in the professor's argument is based on the experience that professors typically provide accurate information. Trust in epistemic authority can also be based on monitoring for the trappings of competence, such as professional titles. This kind of monitoring is more about the symbols of authority than substance \autocite{shieber2015testimony}. 
 
In our study, a source’s gender and ethnicity did not predict participants' credibility judgments. This is in contradiction to some previous work \autocite{armstrong2009blogs, groggel2019race}. In these previous studies, the focus was more on the source's gender and/or ethnicity as such. For example, when examining the gender effects on perceived credibility of blog authors, the content of the blog was kept constant \autocite{armstrong2009blogs}. In our study, the expertise of the source and content of the post were varied, which might explain why gender and ethnicity of the source did not predict participants’ credibility judgments. Thus, it is possible that this variation overrode the effect of the source’s gender and ethnicity.

It is worth noting that the credibility evaluation of social media posts is cognitively demanding as readers must keep several content and source features in their minds when processing and judging the posts. In this light, our results align with the limited capacity model of mediated message processing \autocite{lang2000limited}. The model suggests that due to limited cognitive capacity, people cannot process all aspects of messages they receive, and thus, readers select and process only some features of messages. Furthermore, building on this assumption, the dual processing model of credibility assessment emphasizes that readers’ motivation and ability determine whether and to what degree readers will evaluate the credibility of online information \autocite{metzger2015psychological}. Our credibility evaluation task did not have high personal relevance for the participants, crowdworkers. Thus, crowdworkers’ motivation to deeply consider a broader range of source and content features may have been low, leading to using cognitively and temporally cost-effective evaluation strategies. For example, crowdworkers may have based their credibility evaluations mainly on their prior beliefs and source expertise, neglecting the evaluation of the quality of evidence. We assume that readers’ behavior would likely be somewhat different if they were engaged in seeking information needed to make important health-related decisions (cf. \cite{sperber2010epistemic}). Thus, the results of this study should be interpreted carefully. 

Regarding contextual attributes of the crowdworkers themselves, because there appeared to be no overlap between the self-reported nationalities of study participants in the Prolific and Toloka platforms, we explored the possibility that cultural background (i.e., groups of individuals with nationalities associated with each platform) played a role in predicting the quality of adults’ credibility evaluation practices. Notably, we found that the crowdsourcing platform in which participants completed the credibility evaluation tasks had a significant effect on their performance. Toloka crowdworkers, representing nationalities including Russia, Eastern Europe, the Middle East, Keyna, and India, tended to evaluate the credibility of the inaccurate posts, in particular, higher than Prolific crowdworkers, most of whom lived in the United Kingdom, United States, Canada, and Western and Southern Europe. One reason behind these results might be that adults living in open Westernized countries, like the participants in Prolific, may be exposed to more opportunities to critically question the quality of information sources while participants in Toloka were living in countries like Russia and Turkey with governments that are more likely to repress public forms of criticism and adopt greater restrictions targeting social media use (cf.\cite{tufekci2014social}).Thus, Toloka participants may have less practice in negotiating alternative viewpoints and distinguishing between accurate and inaccurate information in social media posts. 

\subsection{ Implications for Credibility Evaluation Theory and Research}

Findings from this study have implications for theory and research involving the study of credibility evaluation. First, our results affirm the bi-directional model of first- and second-hand evaluation strategies \autocite{barzilai2020dealing} by showing that both content and source related issues were associated with reader’s credibility judgments of social media posts. In addition, by accounting for multiple factors (i.e., variations in content and evidence type, variations in the source’s expertise, gender and ethnicity, and differences in crowdsourcing platform membership) at the same time, rather than assigning participants to predefined conditions involving fewer factors, this study provides a richer and more complex understanding of the order of importance each factor can play in adults’ overall credibility evaluations of social media posts. For example, while we learned that adults’ rely on their prior beliefs and source expertise more than other source features, such as gender and ethnicity, to evaluate overall credibility, we found that evidence quality doesn’t matter as much as we would hope when adult readers consume information about controversial health issues on social media. Our study also provides more conceptual clarity in how variations in these factors can be operationalized and incorporated into the design of authentic, critical online reading tasks for adults.

These findings also complexify bi-directional theories of credibility evaluation \autocites{barzilai2020dealing, richter2017comprehension, stadtler2014content} by hinting at how readers may attend to contextual attributes of the text as a third tier of credibility evaluation \autocite{forzani2022does}. That is, as readers are expected to make inferences about the social and political contexts of health-related social media posts (e.g., unvetted information often paired with testimonials and endorsements by strangers), it is possible that their reasoning may need to extend beyond the use of explicit content and source features to evaluate the quality of unstated warrants in these contexts. Of course, the design of our study does not allow us to explore this possibility. To understand how participants employ first- or second-hand evaluation strategies or if there is a need for additional tiers of strategy use, methods (e.g., think-aloud, open-ended questions) allowing participants to explain their reasoning behind their credibility judgments are needed. Furthermore, more controlled experiments could clarify the nature of the reciprocal relationship between content and source evaluation, or explore the role of additional reader attributes needed to make inferences about information in complicated globally networked social media contexts.

Finally, by broadening our conceptualization of context beyond text and source attributes to explore the effect of one contextual attribute of the participating readers (i.e., crowdsourcing platform membership), we considered the idea that significant differences in credibility evaluations may potentially reflect the different demographic compositions of the participants in the Prolific and Toloka platforms. Unfortunately, because many different nationalities were represented within each platform, it is impossible to know more about how geographic location or nationality might have affected the credibility evaluations of participating crowdworkers in our study. However, the idea of including nationality, and other demographic attributes of readers (i.e., such as gender, ethnicity, and level of education) in future conceptualizations of credibility frameworks seems worthy of additional study.  

Findings from this study also have methodological implications for research in credibility evaluation. Our full factorial design with five independent variables, each having multiple levels, allowed us to study numerous factors within the same experiment, constrained only by the number of unique variations that can be generated. Our methodology also enabled a comprehensive comparison of the influence that each considered factor had on the dependent variable. By employing a cumulative link mixed model (special form of multiple regression) to assess the impact of individual categories within each factor, the resulting coefficients and calculated effect sizes offer a means to rank these factors based on their relative impact. Our study also underscores the potential of applying software to generate and ensure that all unique combinations of social media posts have been created. Key to this approach was the well-established research in combinatorial software testing \autocite{kuhn2009combinatorial}, which involves the automatic generation of unique test cases from a model of all factors and their levels. Crowdsourcing has emerged as a natural methodological choice in modern online combinatorial experiments that demand scalable and on-demand human subject participation. In our study, this approach enables the rapid testing of hundreds of unique variations. Additionally, leveraging multiple platforms promotes cultural and global diversity in samples, as different online human subject pools have distinct participant bases \autocite{douglas2023data}. Achieving full cultural representativeness in study samples will require further efforts of course, and one must recognize the self-selection bias inherent in these subject pools. Nevertheless, platforms like Prolific, Panels \& Samples (by Qualtrics), and Google Surveys are now widely recognized for capturing high-quality human opinions. They are widely adopted by multinational corporations like Meta, institutions such as The World Bank, and behavioral researchers from prestigious institutions, including Stanford and the University of Oxford.

\subsection{Limitations} 

As with all studies, this study has limitations that must be considered when interpreting the results. First, we only investigated two first-hand evaluation strategies (i.e., discourse-based validation and knowledge-based validation), excluding corroboration from our inspection. As corroboration is a vital strategy to understand the consensus in the scientific community \autocite{osborne2022science}, future studies could seek to understand how and in what conditions readers corroborate information when they encounter health information in social media. Furthermore, we did not manipulate the bandwagon features typical of social media messages, such as likes and shares, even though they have been shown to affect the credibility evaluation of such messages \autocite{LIN2016264}. However, we would like to note that the current design already resulted in a large number of different messages (480 unique combinations for accurate posts and 360 unique combinations for inaccurate posts).

Second, we used well-known organizations (e.g., WHO) and non-named experts (e.g., doctor, personal trainer) in the testimonial evidence. Consequently, the strength of the evidence in the testimony and research categories was partly overlapping. In future studies, the second-hand sources referred to in testimonial evidence should be better aligned.

Third, due to the platforms' functionalities, the posts were delivered to the different crowdsourcing platforms slightly differently. However, we attempted to minimize the potential effects by controlling for the platform used in our analysis. However, using two crowdsourcing platforms resulted in a more diverse sample compared to using only one platform. It is worth mentioning that the crowdsourcing platform had an effect on perceived credibility. The crowdworkers of Toloka tended to evaluate the credibility of the inaccurate posts, in particular, higher than the crowdworkers of Prolific. Future studies should consider potential effects when using different crowdsourcing platforms, especially when generalizing findings.

Fourth, using crowd-sourcing platforms for data collection has some downsides. For example, workers may complete the tasks as quickly as possible to maximize their income which is a potential threat to the validity of all studies performed using crowd-sourcing \autocite{gadiraju2015understanding}. We combated this threat by using filters in both crowd-sourcing platforms, that is, recruiting only the top-performing workers for our study.

Fifth, we used a single item to measure the perceived credibility of the social media posts, limiting the estimation of instrument reliability. However, according to \textcite{fuchs2009using}, single-item scale may be advantageous to avoid participants resenting their being asked to respond to similar items, especially when items are administered multiple times.

\section{Conclusions}\label{sec:conclusions}

Our study sheds light on how variations of multiple factors of social media posts (i.e., content, evidence type, source expertise, gender, and ethnicity) can influence credibility evaluation performance. When evaluating the credibility of social media posts, readers can pay attention to the quality of content and trustworthiness of the source of the post. This study shows that readers tend to judge belief-consistent information as more credible than belief-inconsistent information. At worst, this tendency, accompanied by algorithm bias, may strengthen some readers' false beliefs. This, in turn, may accelerate the spread of misinformation in social media. Alarmingly, readers tend to ignore, to some extent, the quality of evidence in their credibility evaluation. This may be due to reluctance to revise one's beliefs, inability to differentiate the quality of different types of evidence, or not valuing research or expert evidence. However, regardless of whether the post was accurate or inaccurate, content posted by an expert was evaluated as more credible than content posted by a layperson. Thus, readers can be more vulnerable to misinformation posted by someone considered an expert. Moreover, our study revealed that different demographic backgrounds of the readers may significantly affect credibility evaluation performance.

Overall, our study suggests a need for further developing the credibility evaluation skills of readers across all ages, not only of children and young people who have typically been the target group of critical reading interventions. An inclusive education system should equip all citizens with adequate credibility evaluation skills. Further, the design and findings of our study answers calls to capture more of the complexities of digital reading tasks associated with texts, readers, and contexts \autocite{coiro2021toward} in addition to calls for more credibility evaluation research that considers the demographic attributes of readers (see \cite{shariff2020review}). Last, our study bridges theory and research efforts in psychology, communication, and media studies with those that explore critical online resource evaluation \autocite{forzani2020three, forzani2022does} in educational contexts.

\printbibliography
\section{Appendix}\label{sec:appendix}

\newpage

\begin{figure*}[t]
\centering
\caption{Pictures of the faces used in our study.}
\includegraphics[width=.80\textwidth]{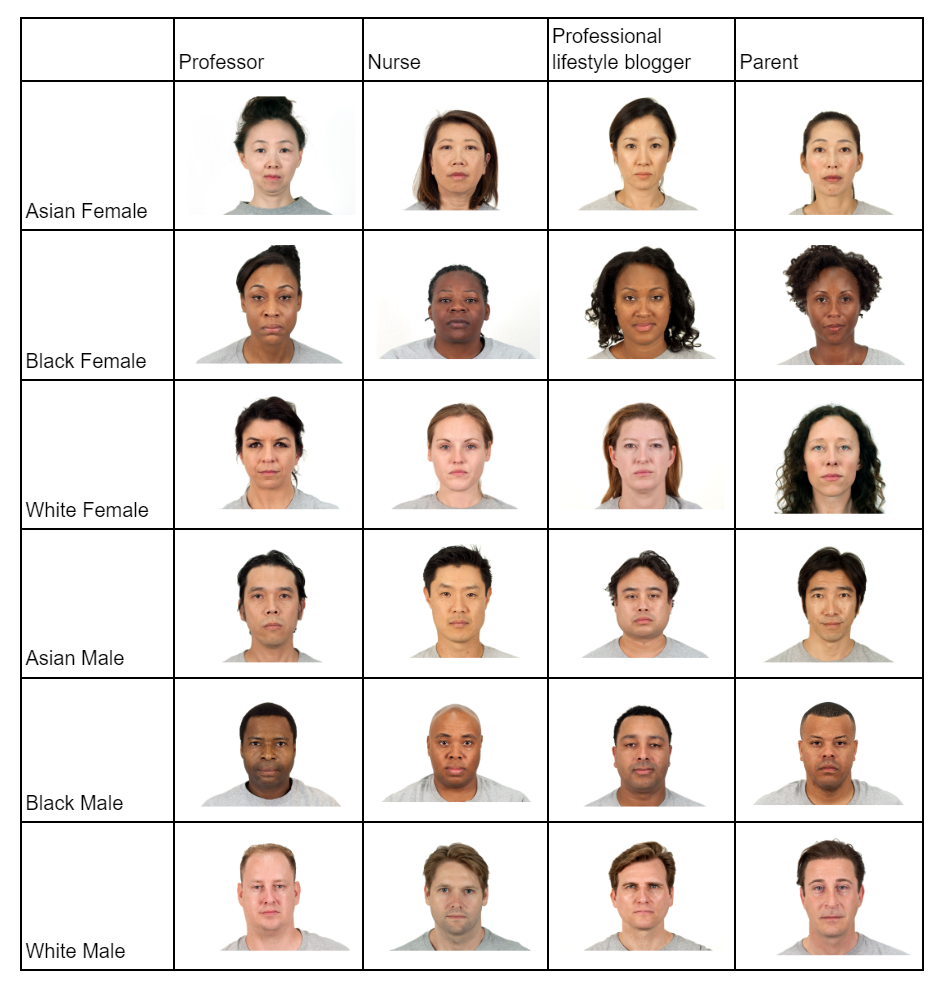}
\label{fig:faces}
\end{figure*}

\begin{table*}[b]
\caption{Second-hand Sources for Posts Presenting Research-based Evidence}
\label{tab:knuutireferences}
\begin{tabular}{|p{7em}|p{34em}|}
  \hline
Topic & Second hand source \\
  \hline
Fish oil & Sinn, N., Milte, C. M., Street, S. J., Buckley, J. D., Coates, A. M., Petkov, J., \& Howe, P. R. (2012). Effects of n-3 fatty acids, EPA v. DHA, on depressive symptoms, quality of life, memory and executive function in older adults with mild cognitive impairment: a 6-month randomised controlled trial. British Journal of Nutrition, 107(11), 1682–1693. \\
\hdashline
Food healthiness & Shan, Z., Guo, Y., Hu, F. B., Liu, L., \& Qi, Q. (2020). Association of low-carbohydrate and low-fat diets with mortality among US adults. JAMA Internal Medicine, 180(4), 513–523.\\
\hdashline
Red processed meat  & Zhong, V. W., Van Horn, L., Greenland, P., Carnethon, M. R., Ning, H., Wilkins, J. T., ... \& Allen, N. B. (2020). Associations of processed meat, unprocessed red meat, poultry, or fish intake with incident cardiovascular disease and all-cause mortality. JAMA Internal Medicine, 180(4), 503–512.\\
\hdashline
Vaccines & "Gołoś, A., \& Lutyńska, A. (2015). Aluminium-adjuvanted vaccines–a review of the current state of knowledge. Przegl Epidemiol, 69(4), 731–734.\\
\hdashline
Vitamin D  & Manson, J. E., Cook, N. R., Lee, I. M., Christen, W., Bassuk, S. S., Mora, S., ... \& Buring, J. E. (2019). Vitamin D supplements and prevention of cancer and cardiovascular disease. New England Journal of Medicine, 380(1), 33–44. \\
\hline
\end{tabular}
\end{table*}









%



\end{document}